\documentclass[aps,prl,showpacs,twocolumn,amsmath,amssymb,superscriptaddress,footinbib]{revtex4}

\usepackage[english]{babel}
\usepackage{latexsym}

\usepackage{color}
\usepackage{graphics}
\usepackage{subfigure}
\usepackage{epsfig}
\usepackage{amsfonts}
\usepackage{amssymb}
\usepackage{amsmath}
\usepackage{bbm}

\newcommand{\ie}{{\it i.e.}}

\newcommand{\via}{{\it via}~}

\newcommand{\n}{\mathbf{n}}

\newcommand{\ket}[1]{\ensuremath{|#1\rangle}}

\newcommand{\ew}[1]{\ensuremath{\langle #1\rangle}}
\begin{document}

\title{Ultracold atomic Bose and Fermi spinor gases in optical lattices}

\author{K. Eckert}
\affiliation{Departament F\' isica, Grup F\' isica Te\'orica, 
Universitat Aut\'onoma de Barcelona, E-08193 Bellaterra, Spain.}
\author{\L. Zawitkowski}
\affiliation{Centrum Fizyki Teoretycznej, Polska Akademia Nauk, Warszawa  02668, Poland.}
\author{M.J. Leskinen}
\affiliation{Nanoscience Center, Department of Physics, PO Box 35, University of Jyv\"askyl\"a, Finland}
\author{A. Sanpera}
\affiliation{ICREA: Instituci\'o Catalana de Recerca i Estudis Avan\c cats.} 
\affiliation{Departament F\' isica, Grup F\' isica Te\'orica, 
Universitat Aut\'onoma de Barcelona, E-08193 Bellaterra, Spain.}
\author{M.~Lewenstein}
\affiliation{ICREA: Instituci\'o Catalana de Recerca i Estudis Avan\c cats.} 
\affiliation{ICFO--Institut de Ci\`encies Fot\`oniques,
E-08860 Castelldefels, Barcelona, Spain.}
\affiliation{Institut f\"ur Theoretische Physik, Universit\"at Hannover, D-30167 Hannover, Germany.}

\date{\today}

\begin{abstract}
We investigate magnetic properties of Mott-insulating phases 
of ultracold Bose and Fermi spinor gases in optical lattices. We consider 
in particular the $F=2$ Bose gas, and the $F=3/2$ and $F=5/2$ Fermi gases. 
We derive effective spin Hamiltonians for one and two atoms per site and
discuss the possibilities of manipulating the magnetic properties of the 
system using optical Feshbach resonances. We discuss low temperature quantum phases of a
$^{87}$Rb gas in the $F=2$ hyperfine state, as well as  possible realizations of high spin Fermi gases with
either $^6$Li or $^{132}$Cs atoms in the $F=3/2$ state, and with $^{173}$Yb atoms in the $F=5/2$ state.
\end{abstract}


\maketitle

\section{I. Introduction}


\paragraph{\bf Ultracold spinor gases}
The seminal theory papers by T.-L. Ho \cite{Ho98}, and T. Ohmi and K. Machida \cite{machida} on spinor $F=1$ Bose-Einstein Condensates (BECs)
as well as the experiments performed by the MIT group 
on optically trapped $F=1$ Sodium condensates \cite{Sten98},
have brought a new perspective to the study of magnetic systems 
using ultracold atomic gases. 
Interactions in bosonic systems with 
spin degrees of freedom host a wide variety of exotic phases 
at zero temperature and a dynamics clearly differentiated from the one
displayed by scalar condensates. 
Recent studies involving both, $F=1$ and $F=2$ Rubidium atoms, 
have focused, for instance, on the rich dynamics of spinor Bose condensates  
\cite{sekkscabsPRL2004,Chang04,Barrett01,Kuwamoto04}. 
Ground-state and dynamical properties of $F=3$ Bose condensates have also been discussed in
connection to ongoing experiments with Chromium atoms \cite{ho2005}.

\paragraph{\bf Spinor Bose gases in optical lattices} 
Several experimental groups, who until now achieved spin $F=1$ and $F=2$ condensates in optical traps, have already started, or are indeed planning, experiments with bosonic spinor  lattice gases \cite{privada}.   
Confining spinor BECs in optical lattices 
offers a unique opportunity to study magnetic properties of matter, 
as a large range of tunable parameters exists, which are not accessible in solid state systems (for a review see \cite{ml}). For instance, disorder, or pseudo-disorder, can be created in a highly controlled way in optical lattices
\cite{Damski,Roth}. This broadens enormously the variety  of magnetic systems which can be "mimicked"  with ultracold gases. Spinor gases also offer the possibility of measuring quantum fluctuations of the total magnetization of the system employing quantum Faraday effect \cite{eckert:2006,cherng:2006}. 
Another particularly interesting application of 
ultracold spinor gases in optical lattices is the  
engineering of strongly correlated states, and processing of quantum information. The first steps toward these goals have been achieved by A. Widera {\it et al.} \cite{widera}, who investigated a method of
measuring the coherent  collisional spin dynamics in a lattice. This method has allowed for a very precise determination of the $^{87}$Rb scattering lengths for $F=1$ and $F=2$ \cite{blochas}.  

\paragraph{\bf $\mathbf{F=1}$ gases in optical lattices}
Studies of  $F=1$ systems have already been carried 
out by Demler's group \cite{Demler}. They have derived an approximate phase diagram for the case of antiferromagnetic interactions of $^{23}$Na. As in the standard Bose-Hubbard model, an $F=1$ spinor gas undergoes superfluid (SF) to Mott insulator (MI) transition as tunneling is decreased. In the antiferromagnetic case in 2D and 3D, the SF phase is {\it polar}, and so are the Mott states with an odd number $N$ of atoms per site (those states are also termed as {\it nematic}). In the case of even $N$, for small tunneling the Mott states are singlets, and for moderate tunneling there occurs a first order transition to the nematic state.

In 1D there is furthermore the possibility of a dimerized state, as in the Majumdar-Ghosh model 
\cite{auerbach}. This possibility was studied by Yip \cite{yip1}, who derived an effective spin Hamiltonian for the MI state with $N=1$. Using a variational ansatz interpolating between dimer and nematic states indicated that in a wide range of parameters the spinor $^{23}$Na lattice gas should have a dimerized ground state in 1D, 2D, and 3D. This is a very interesting result, since so far dimer states have not been observed in experiments. This result has been supported by rigorous studies in Ref. \cite{yip2}. It was shown that under the same effective Hamiltonian, for a system with an even number of sites the ground state  has total spin $S_{tot}=0$, while the first excited state has $S_{tot}=2$. Yip's results were recently confirmed by Rizzi {\it et al.} \cite{rizzi}, who numerically studied the SF -- MI transition in the $F=1$ Bose-Hubbard model in 1D. They found that in the low tunneling regime of the first MI lobe (where the effective spin model of Ref. \cite{Demler} works) the system is always in a dimerized state. Similar results were obtained by Porras  {\it et al.} \cite{diego}. Thus, strictly speaking, nematic order seems to be absent in 1D in the thermodynamic limit. However, susceptibility to nematic ordering grows close to the border of the ferromagnetic phases, indicating that it may persist in finite systems.   

Another interesting aspect, namely the possibility of controlling the order of the SF -- MI transition by using appropriately polarized (lin--$\theta$-lin) laser fields to form the optical lattice was investigated in 
Refs. \cite{graham,Kimura}. Such a laser configuration couples the states with $m_F=\pm 1$, so that the system becomes effectively two--component.

\paragraph{\bf $\mathbf{F=2}$ gases in optical lattices}
The mean field states of spinor $F=2$ gases have been for the first time investigated in Refs.\ \cite{cyhPRA2000,kuPRL2000,ukPRA2002}. It is worth noticing that mean field states are also valid for MI states with one atom per lattice site, provided all atoms are described by the same single-particle wave function attached to a given site. Refs. \cite{kuPRL2000,ukPRA2002} go one step further, and apart from 
the mean field theory consider also the extreme case of quenched (immobile) $F=2$ bosons in an optical lattice. In other words, these articles characterize possible on--site states for $N$ bosons with total spin $S$ in the absence of tunneling. 

After submission of the first version of this paper, Barnett, Turner, and Demler presented  a beautiful and complete classification of the mean field phases for arbitrary $F$, based on 19th century method by F.~Klein of solving quintic polynomials by the analysis of rotations of regular icosahedra \cite{barnett}. We discuss their results in more details in the following. Also, very recently the effective spin Hamiltonians (in the first MI lobe), and quantum insulating phases  of $F=2$ bosons have been studied by Zhou amd Semenoff \cite{zhoulast}, applying the variational principle to product (Guztwiller ansatz, cf. \cite{jPRL1998}), dimer, and trimer states. Their results concerning efffective spin Hamiltonians agree with ours.   

\paragraph{\bf Spinor Fermi gases in optical lattices} 
Obviously, there is an enormous interest also in Fermi gases, and in particular in spinor Fermi gases in optical lattices. The first reason is, of course, that such systems could realize a perfect quantum simulator of the fermionic Hubbard model, and thus shine some light on the problem of high $T_c$ superconductivity. For spin $F=1/2$ this has been proposed in Ref.~\cite{hofstetter}, and recently considered with three--component fermions in Ref.~\cite{paananen}. 

Liu {\it et al.} \cite{wilczek} proposed to use fermions with high $F$ to realize spin-dependent Hubbard models, in which hopping parameters are spin-dependent. Such models lead to exotic kinds of superfluidity, such as to a phase in which SF and normal component coexist at zero temperature. 

W.~Hofstetter and collaborators have written a series of papers, reviewed in \cite{hofstetter2}, on fermionic atoms with $SU(N)$ symmetry in optical lattices.
Such systems also  have exotic superfluid and flavor-ordered ground states, and exhibit very rich behavior in the presence of disorder.   

It is, of course, inevitable to ask which atoms can be used to realize high--$F$ fermonic spinor gases in optical lattices. The most commonly used alkali $^6$Li has hyperfine manifolds with $F=1/2$ and $F=3/2$. The latter is, obviously, subjected to
two--body losses, but as in the case of the $F=2$ manifold of Rubidium, one can expect reasonably long life time in the lattice (especially in MI states with $N=1$). Another commonly used fermion is a heavy alkali $^{40}$K, which has manifolds $F=7/2$ and $F=9/2$. These fermions are particularly useful for spin-dependent Hubbard models 
\cite{wilczek}.

There are several atoms whose lowest hyperfine manifold has $F=3/2$, \ie, in those ground states two body losses can be avoided: $^9$Be, $^{132}$Cs, or $^{135}$Ba, but so far only the bosonic Cesium BEC has been achieved \cite{rudi}. On other hand, recently a BEC of $^{174}$Yb atoms \cite{ybbec}, as well as degenerate gas of $^{173}$Yb fermions with $F=5/2$, has been realized. Finally, fermionic Chromium has 4 hyperfine manifolds with
$F=9/2,7/2,5/2,3/2$, in the ascending order of energies, and after achieving BEC of the bosonic Chromium \cite{pfau}, the prospect for achieving ultracold degenerate Fermi gases are very good.

\paragraph{\bf $\mathbf{F=3/2}$ and $\mathbf{F=5/2}$ Fermi gases in optical lattices}
Recently, there has been a lot of progress in understanding the special properties of $F=3/2$ and $F=5/2$
Fermi gases. C. Wu {\it et al.} \cite{wu1} realized that the spin-3/2 fermion models with contact interactions have a
generic $SO(5)$  symmetry without any fine-tuning of parameters, and employed this fact to propose a quantum Monte Carlo algorithm free of the sign-problem. They have found novel competing orders \cite{wu2} in spin-3/2 cold atomic
systems in one-dimensional optical traps and lattices. In particular,
the quartetting phase, a four-fermion counterpart of Cooper pairing,
exists in a large portion of the phase diagram. Recently, these authors studied the $s$-wave quintet 
Cooper pairing phase (having $S_{tot}=2$) in spin-$3/2$  atomic gases, and identified various novel features which
do not appear in the spin-1/2 counterpart. For instance, a single quantum vortex was
shown to be energetically less stable than a pair of half-quantum vortices.

The bosonization approach was applied to 1D systems with $F=3/2,\,5/2,\,\ldots$ by Lecheminant {\it et al.} \cite{lecheminant}. They used a somewhat simpler model with a spin independent coupling $U$ and a coupling in the singlet channel $V$, and studied the phase diagram in the $U-V$ plane. They found 3 phases: a spin density wave, an atomic density wave (which may crossover to a molecular superfluid), and a BCS superfluid (that may crossover to a molecular density wave). They have 
also classified Mott phases at commensurate $1/(F+1/2)$ fillings.  Finally, the $F=3/2$ model has been solved analytically using Bethe ansatz and identifying the effective low energy field theory describing the system to be that of a deformed Gross--Neveu model with either $O(7)\times {\mathbb Z}_2$ symmetry at half--filling, or 
$U(1)\times O(5)\times {\mathbb Z}_2$ symmetry otherwise \cite{controzzi}. 
An overview of hidden symmetries and competing orders in spin 3/2 gases is
presented in the excellent paper of C. Wu \cite{wureview}. 
In a very recent preprint, Tu {\it et al.} \cite{tu} have studied spin quadrupole ordering in spin-$3/2$ gases and derived effective spin Hamiltonians for the first and second MI lobe. Our results agree with theirs and provide complementary approach and discussion.

\paragraph{\bf Goals of the paper}
Concerning the Bose gases, the main goal of the present paper is to make predictions for the planned experiments with ultracold $F=2$ $^{87}$Rb atoms in the strongly correlated regime. We will thus concentrate on characterizing possible MI states in the limit of weak tunneling. In this sense we will generalize the results of Refs. 
\cite{kuPRL2000,ukPRA2002}.  We will disregard here the Zeeman effect by assuming sufficiently efficient magnetic shielding. Also, we will only consider the MI states with $N=1$ or $N=2$ atoms per site. Higher atom numbers will inevitably lead to 3--body losses, and will be thus much more difficult to realize experimentally.

We will also analyze the possibility of exploring parameters of the systems by modifying atomic scattering lengths. This cannot be done using the standard Feshbach resonances (cf. \cite{timmermans}), since we assume zero magnetic field. Instead, one has to use the method of optical Feshbach resonances, proposed by Fedichev {\it et al.} \cite{fedichev}. 

Our main goal will be to derive effective spin models in the experimentally relevant regimes to study the ground states of the systems, and in particular to identify those instances when exact solutions are available, either in the form of product (mean field) states, or the so called matrix product states (MPS) \cite{werner}, similarly as it happens in the famous AKLT model \cite{aklt}.

Similar goals concern Fermi gases, although there much less is known about values of scattering lengths, the possibility of optical Feshbach resonances, etc. We will take our liberty here to explore larger regions of parameters, assuming optimistically that they will become feasible experimentally at some point.

\paragraph{\bf Plan of the paper} 
In Sec.~II we discuss the Bose-Hubbard Hamiltonian
of $F=2$ bosons in an optical lattice and its ground states when 
atoms are quenched at fixed lattice sites. In Sec.~III
we derive an effective spin Hamiltonian for this system. 
In Sec.~IV we investigate which types of ground states could be achieved, 
assuming a (limited) experimental control over the spin-dependent scattering
lengths. In Secs.~V and VI we analyze the Fermi-Hubbard and the effective Hamiltonian for 
$F=3/2$, and in Secs. VII and VIII we perform a similiar analysis for $F=5/2$.
We conclude in Sec.~IX. Appendix A contains a detailed analysis of possibilities of optical
manipulations of scattering lengths of  $^{87}$Rb atoms in the $F=2$ hyperfine manifold.
Appendix B gives a short overview of MPS and PEPS methods.

\section{II. $F=2$ Spinor Bose-Hubbard Hamiltonian}

\paragraph{\bf The system} We consider $F=2$ atoms at low temperatures 
confined in a deep optical lattice so that the system  
is well described by a Bose-Hubbard Hamiltonian \cite{jPRL1998}.
We assume atoms to interact {\it via} a 
a zero-range potential. 
The total angular momentum of two colliding identical bosons 
is restricted to even values due to Bose symmetry, 
so that the $s$-wave interaction between two spin-2 particles can be written as
$\hat{V}=\bar g_0 \hat{P}_0 + \bar g_2 \hat{P}_2 + \bar g_4 \hat{P}_4$, 
where $\hat{P}_S$ ($S=0,2,4$) is the projector onto the 
subspace with total spin $S$. 
The interaction strengths, $\bar g_S$, depend on the total 
spin of the two colliding particles. For the different channels they are given by the scattering lengths $a_S$ through
$\bar g_S=\frac{4\pi\hbar^2 a_S}{m}$.

To better understand ground state properties of the Hamiltonian it is convenient to
express the interaction potential $\hat{V}$ in terms of spin operators.  
Making use of the identities
$\hat{I}=\hat{P}_0+\hat{P}_2+\hat{P}_4$ and 
$\hat{\textbf{\textit{F}}}_1\cdot \hat{\textbf{\textit{F}}}_2=-6\hat{P}_0-3\hat{P}_2+4\hat{P}_4$,
where $\hat{\textbf{\textit{F}}}_{i}$ corresponds to the spin operator of
particle $i$, we find
$\hat{V}=\bar c_0 \hat{I}+ \bar c_1 \hat{\textbf{\textit{F}}}_1\cdot \hat{\textbf{\textit{F}}}_2 + \bar c_2 \hat{P}_0$. Here
$\bar c_0=(3 \bar g_4+4 \bar g_2)/7$, $\bar c_1=(\bar g_4-\bar g_2)/7$, and $\bar c_2=(3 \bar g_4-10 \bar g_2+7 \bar g_0)/7$ 
\cite{kuPRL2000}. Following \cite{ukPRA2002},  
$\hat{P}_0$ can also be expressed in terms of  "singlet pair" 
creation and annihilation operators
$\hat{S}_+=\hat{a}^{\dagger}_0 \hat{a}^{\dagger}_0/2-\hat{a}^{\dagger}_1 \hat{a}^{\dagger}_{-1}+ 
\hat{a}^{\dagger}_2 \hat{a}^{\dagger}_{-2}$, $\hat{S}_-=\hat{S}^{\dagger}_+$, where
$\hat{a}_{\sigma}^{\dagger}$ ($\hat{a}_{\sigma}$) creates (annihilates) a particle with spin projection $\sigma$.
The operator $\hat{S}_+$ applied on the vacuum creates, except for
normalization, two bosons in a spin singlet state.
Such a pair does not represent a composite boson, as $\hat{S}_+$ and $\hat{S}_-$ 
do not satisfy Bose commutation relations.
$\hat{P}_0$ can now be written in second quantization as $\hat{P}_0=2 \hat{S}_+ \hat{S}_-/5$. Its eigenvalues are $N_S (2N-2N_S+3)/5$, where the quantum number $N_S$ 
denotes the number of spin-singlet pairs and $N$ the total number of bosons \cite{ukPRA2002}. 

\paragraph{\bf Bose-Hubbard Hamiltonian} The Bose-Hubbard Hamiltonian (BHH) for spin $F=2$ can be written as
\begin{equation}
\hat{H} = -t\sum_{<ij>,\sigma}(\hat{a}^{\dagger}_{\sigma i} \hat{a}_{\sigma j}+\hat{a}^{\dagger}_{\sigma j} \hat{a}_{\sigma i})
+\sum_{i,S}g_s \hat{P}_{Si},\label{latwyham}
\end{equation}
or alternatively
\begin{eqnarray}
\hat{H}&=&-t\sum_{<ij>,\sigma}(\hat{a}^{\dagger}_{\sigma i} \hat{a}_{\sigma j}+\hat{a}^{\dagger}_{\sigma j} \hat{a}_{\sigma i})
+\frac{c_0}{2}\ \sum_{i}\hat{N}_i(\hat{N}_i-1)\nonumber \\
&&+\frac{c_1}{2} \sum_{i}:\hat{\textbf{\textit{F}}}_i
\cdot \hat{\textbf{\textit{F}}}_i:
+\frac{2 c_2}{5}\sum_i \hat{S}_{+i} \hat{S}_{-i},\label{eqn:hubham}
\end{eqnarray}
where $\hat{a}_{\sigma i}$ annihilates a particle in a hyperfine state $m_F=\sigma$ at site $i$, 
$\hat{N_i}=\sum_{\sigma}\hat{a}^{\dagger}_{\sigma i} \hat{a}_{\sigma i}$ is the number of particles at site $i$, 
$\hat{\textbf{\textit{F}}}_i=\sum_{\sigma \sigma^{\prime}} \hat{a}^{\dagger}_{\sigma i} 
\textbf{\textit{T}}_{\sigma \sigma^{\prime}} \hat{a}_{\sigma^{\prime} i}$ is the spin operator at site $i$ 
($\textbf{\textit{T}}_{\sigma \sigma^{\prime}}$ being the usual spin matrices for a spin-2 particle), and $:\hat{X}:$ denotes normal ordering of the operator $\hat{X}$.
The coefficients $c_i=\bar c_i\int d^3x\ |w(x)|^4$, and $g_S=\bar g_S\int d^3x\ |w(x)|^4$, where $w(x)$ is the Wannier function centered at $x=0$. 
The first two terms in the Hamiltonian represent tunneling between nearest-neighbor sites and Hubbard repulsion between atoms on the same site, respectively, 
as in the standard Bose-Hubbard model.
The two remaining terms represent the energy associated with spin configurations within lattice sites. 
The ratios between the various interactions, $c_1/c_0$ and $c_2/c_0$, are fixed by the scattering lengths 
($c_1/c_0=\bar c_1/\bar c_0$ and $c_2/c_0=\bar c_2/\bar c_0$).
We assume here that the scattering lengths are such that the Hamiltonian (\ref{eqn:hubham}) is stable with respect
to collapse. Stability requires the $g_S\ge 0$ for all $S$, which is the case, e.g.,
for $^{87}$Rb. The ratio $t/c_0$ between tunneling and Hubbard repulsion can be tuned by changing the lattice
parameters \cite{jPRL1998}.
When $t\ll c_0$, the system is in a Mott-insulating phase in which atoms are quenched at fixed lattice sites.
We consider here the case when tunneling is sufficiently weak compared to Hubbard repulsion so that it can be treated as a perturbation with $t/c_0$ being a small parameter.
We study systems with one and two particles per site,
which are the most interesting cases as they do not suffer from three-body losses.

\paragraph{\bf On--site Hamiltonian} To zeroth order, the Hamiltonian is a sum of independent single-site Hamiltonians (we omit the index $i$):
\begin{equation}\label{H_0}
\hat{H}_{0}=\frac{c_0}{2}\ \hat{N}(\hat{N}-1)+\frac{c_1}{2}\,:\hat{\textbf{\textit{F}}}\cdot \hat{\textbf{\textit{F}}}: 
+ \frac{2 c_2}{5} \hat{S}_+ \hat{S}_-.
\end{equation}
Exact eigenstates of this Hamiltonian have been obtained in Ref. \cite{ukPRA2002}. 
Since $\hat{S}_{\pm}$ commute with the total spin operator, the energy eigenstates can be labeled with four quantum numbers as $|N,N_S,\bar F\rangle_{m_{\bar F}}$, where $N$ is the number of particles per site, $N_S$  the number of spin-singlet pairs, and
$\bar F$ is the total on-site spin.
The eigenstates have a $2\bar F+1$-fold degeneracy associated with the quantum number $m_{\bar F}$. 
Their  energies are:
\begin{eqnarray}\label{energies}
E&=&\frac{c_0}{2} N(N-1) + \frac{c_1}{2}(\bar F(\bar F+1)-6N)+\nonumber \\ 
&&\frac{c_2}{5}N_S(2N-2N_S+3).
\end{eqnarray}
In general there is an additional degeneracy which, however, manifest itself only for states with larger number $N$
of particles per site than considered in this paper. 
Single particle states are 
denoted as $|1,0,2\rangle_{m_F}=\hat{a}^{\dagger}_{m_F}|\Omega\rangle$, being 
$|\Omega\rangle$ the vacuum.
Two and three particle states with maximal spin projection $m_{\bar F}$ are \cite{ukPRA2002}: 
\begin{flushleft}
$|2,1,0\rangle_0=\frac{1}{\sqrt{10}}[\hat{a}^{\dagger}_0 \hat{a}^{\dagger}_0-2\hat{a}^{\dagger}_1 \hat{a}^{\dagger}_{-1}
+2\hat{a}^{\dagger}_2 \hat{a}^{\dagger}_{-2}]|\Omega\rangle $,
$|2,0,2\rangle_2=\frac{1}{\sqrt{14}}[2\sqrt{2}\hat{a}^{\dagger}_2 \hat{a}^{\dagger}_0-\sqrt{3}\hat{a}^{\dagger}_1 \hat{a}^{\dagger}_1] |\Omega\rangle $,
$|2,0,4\rangle_4=\frac{1}{\sqrt{2}}\hat{a}^{\dagger}_2 \hat{a}^{\dagger}_2|\Omega\rangle $,
$|3,0,0\rangle_0=\frac{1}{\sqrt{210}}[\sqrt{2}\hat{a}^{\dagger}_0\hat{a}^{\dagger}_0\hat{a}^{\dagger}_0
-3\sqrt{2}\hat{a}^{\dagger}_1\hat{a}^{\dagger}_0 \hat{a}^{\dagger}_{-1}
+3\sqrt{3}\hat{a}^{\dagger}_1\hat{a}^{\dagger}_1\hat{a}^{\dagger}_{-2}
+3\sqrt{3}\hat{a}^{\dagger}_2\hat{a}^{\dagger}_{-1}\hat{a}^{\dagger}_{-1}
-6\sqrt{2}\hat{a}^{\dagger}_2\hat{a}^{\dagger}_0\hat{a}^{\dagger}_{-2}]|\Omega\rangle $,
$|3,1,2\rangle_2=\frac{1}{\sqrt{14}}\ [\hat{a}^{\dagger}_2\hat{a}^{\dagger}_0 \hat{a}^{\dagger}_0
-2\hat{a}^{\dagger}_2\hat{a}^{\dagger}_1 \hat{a}^{\dagger}_{-1}
+2\hat{a}^{\dagger}_2\hat{a}^{\dagger}_2 \hat{a}^{\dagger}_{-2}]|\Omega\rangle $,
$|3,0,3\rangle_3=\frac{1}{\sqrt{20}}[\hat{a}^{\dagger}_1\hat{a}^{\dagger}_1\hat{a}^{\dagger}_1
-\sqrt{6}\hat{a}^{\dagger}_2\hat{a}^{\dagger}_1 \hat{a}^{\dagger}_0
+2\hat{a}^{\dagger}_2\hat{a}^{\dagger}_2\hat{a}^{\dagger}_{-1}]|\Omega\rangle $,
$|3,0,4\rangle_4=\frac{1}{\sqrt{22}}[2\sqrt{2}\hat{a}^{\dagger}_2\hat{a}^{\dagger}_2 \hat{a}^{\dagger}_0-
\sqrt{3}\hat{a}^{\dagger}_2\hat{a}^{\dagger}_1 \hat{a}^{\dagger}_1] |\Omega\rangle $,
$|3,0,6\rangle_6=\frac{1}{\sqrt{6}}\hat{a}^{\dagger}_2\hat{a}^{\dagger}_2\hat{a}^{\dagger}_2 |\Omega\rangle.$
\end{flushleft}
States with lower $m_{\bar F}$ can be obtained by acting with the 
spin lowering operator on the above listed states. 

\paragraph{\bf Phases at t=0.} It is easy to check which phases will be realized in the limit of vanishing tunneling for a given chemical potential. For $\mu<0$ the state with no atoms has the smallest  (Gibbs potential) energy $G=E-\mu N=0$. For $\mu\ge 0$ we enter the phase with one atom for site $|1,0,2\rangle$ with $G=-\mu$. As we increase $\mu$ further, we enter one of the phases with two atoms per site, namely the one that correspond to the smallest $g_S$ and $G=g_S-2\mu$: i) $|2,1,0\rangle$ if $g_0\le g_2,g_4$, ii)  $|2,0,2\rangle$ if $g_2\le g_0,g_4$, and iii) $|2,0,4\rangle$ if $g_4\le g_2,g_0$.

\section{III. Effective Hamiltonian (bosonic case)}

To derive the effective Hamiltonian to second order in $t$ 
we consider the two-site problem. The tunneling Hamiltonian 
$H_t=-t \sum_{\sigma,<ij>} (\hat{a}^{\dagger}_{\sigma i}\hat{a}_{\sigma j}+\hat{a}^{\dagger}_{\sigma j}\hat{a}_{\sigma i})$
conserves both, the total spin $S$ and the projection of the total spin $m_S$ \cite{Demler}.
Thus, to second order, the shift of the energy of the two-site ground state $|g,S\rangle$ with total spin $S$
is given by
\begin{equation}
\epsilon_S=-\sum_{\nu} \frac{|\langle \nu|\hat H_t|g,S\rangle|^2}{E_{\nu}-E_{g,S}},\label{eq:eps_s}
\end{equation}
where $\nu$ labels the (virtual) intermediate states and $E_{\nu},\,E_{g,S}$ 
denote the unperturbed energies of the two-site states $|\nu\rangle$, $|g,S\rangle$
(which are non-degenerate apart from the $m_F$ degeneracy). The dependence of the  
energy shifts on the total spin of the two sites 
introduces nearest-neighbor spin-spin interactions in the lattice.
It is sufficient to evaluate these shifts for only one value of the projection $m_S$
of the total spin. 
This is because tunneling cannot mix states with different $m_S$ and overlaps 
$|\langle \nu|H_t|g,S\rangle|$ are rotationally invariant. 
To simplify the calculations we choose always the highest possible value 
of $m_S$. 

\subsection{A. One atom per site}

\paragraph{\bf Pair Hamiltonian} For a single particle per site, \ie, $|1,0,2\rangle^{(i)}\otimes|1,0,2\rangle^{(j)}$,
only 6 intermediate states are possible:
\begin{equation}
\begin{split}
|\Omega\rangle^{(i)}\otimes|2,1,0\rangle^{(j)}&\,\,\text{and}\,\, i\leftrightarrow j\,\,\,\,(S=0),\\
|\Omega\rangle^{(i)}\otimes|2,0,2\rangle^{(j)}&\,\,\text{and}\,\, i\leftrightarrow j\,\,\,\,(S=2),\\
|\Omega\rangle^{(i)}\otimes|2,0,4\rangle^{(j)}&\,\,\text{and}\,\, i\leftrightarrow j\,\,\,\,(S=4).\\
\end{split}\nonumber
\end{equation}
The corresponding energy shifts are
\begin{equation}
\epsilon_S=-\frac{4 t^2}{g_S},
\end{equation}
or, written in terms of the $c_i$'s,
\begin{equation}
\epsilon_0=-\frac{4 t^2}{c_0+c_2-6c_1},\,\,
\epsilon_2=-\frac{4 t^2}{c_0-3c_1},\,\,
\epsilon_4=-\frac{4 t^2}{c_0+4c_1}.\nonumber
\end{equation}
In this case, the overlap  $\langle \nu|H_t|g,S\rangle$ is the same 
for all virtual tunneling states $|\nu\rangle$. 
Therefore, $\epsilon_S$ depends only on the difference of scattering lengths 
$a_S$. This suggests that control and engineering of the 
magnetic properties of the system could be achieved 
using optical Feshbach resonances \cite{fedichev, grimm}.
One can in principle also use magnetic fields to control the scattering
properties, but that would inevitably lead to (linear and/or  quadratic)
Zeeman effects, which would change the structure of $\hat{H}_0$ (Eq. (\ref{H_0})).
Corresponding effects will be discussed elsewhere.

The effective spin-spin Hamiltonian in second order reads 
\begin{equation}\label{1particleH}
\hat{H}_I^{(ij)}=\epsilon_0 \hat P_0^{(ij)}+\epsilon_2 \hat P_2^{(ij)}+\epsilon_4 \hat P_4^{(ij)}.
\end{equation}
For $^{87}$Rb and using the scattering lengths at zero magnetic field,
the energy shifts are 
$\epsilon_0=- (4t^2/g_0)$, $\epsilon_2=-0.962(4t^2/g_0)$, and 
$\epsilon_4=-0.906(4t^2/g_0)$. Thus, as $\epsilon_0$ is smallest, $^{87}$Rb should experience 
antiferromagnetic behavior in the ground state (cf. \cite{cyhPRA2000}).
For $^{85}$Rb, the scattering lengths are negative in the absence of a magnetic field. Thus
Hamiltonian (\ref{eqn:hubham}) is unstable with respect to collapse. One can, however, use
optical Feshbach resonances (similarly as has been demonstrated using magnetic Feshbach resonances
\cite{cornish}) to achieve $g_S>0$ for all $S$. In this case we expect the physics of
the $^{85}$Rb lattice spinor gas to be similar to the case of $^{87}$Rb.
Despite the fact that control of magnetic properties is possible using 
optical Feshbach resonances, observation of the ground states 
requires, in turn, very low temperatures to resolve accurately the different
energy shifts.

\paragraph{\bf Lattice Hamiltonian} The Hamiltonian (\ref{1particleH}) can be easily generalized to 
the whole lattice,
$\hat{H}= \sum_i \hat{H}_{0,i}+\sum_{<ij>} \hat{H}^{(ij)}_I$.
It can also be transformed into a polynomial of 
fourth order in the Heisenberg interaction 
$\hat{\textbf{F}}_i \cdot\hat{\textbf{F}}_j$:
\begin{eqnarray}
\hat{H}= \sum_i \hat{H}_{0,i}&+&
\sum_{<ij>} \left[
\frac{39\epsilon_0-80\epsilon_2}{51}( \hat{\textbf{F}}_i \cdot \hat{\textbf{F}}_j)\right.
\nonumber \\
&+&\frac{9\epsilon_0-8\epsilon_2}{102}( \hat{\textbf{F}}_i \cdot \hat{\textbf{F}}_j)^2
\nonumber \\
&+&\left(-\frac{7\epsilon_0}{204}+\frac{10\epsilon_2}{204}+\frac{\epsilon_4}{72}\right)(\hat{\textbf{F}}_i \cdot \hat{\textbf{F}}_j)^3
\nonumber \\
&+&\left.\frac{7\epsilon_0+10\epsilon_4}{1020}( \hat{\textbf{F}}_i \cdot \hat{\textbf{F}}_j)^4\right].
\end{eqnarray}
Here up to four powers of $\hat{\textbf{F}}_i \cdot \hat{\textbf{F}}_j$ appear due to the fact that 
channels 1 and 3 between different sites are not forbidden.

\subsection{B. Two atoms per site}

\paragraph{\bf Accessible single site states} We now derive the effective Hamiltonian for the case in which the ground state
corresponds to two particles per site and proceed as before by considering the two-site problem.
To calculate the energy shifts, we find first the ground states 
of the unperturbed Hamiltonian $\hat{H_0}$, which are easily 
determined from Eq. (\ref{energies}). The ground state corresponds to
\begin{itemize}
\item $|2,1,0\rangle$ if  $g_0<g_2,g_4$, \ie, for $c_1<0$, $c_2< 10c_1$ and $c_1>0$, $c_2<3c_1$. This singlet state is 
the single-site ground state of $^{87}$Rb for unmodified values of scattering lengths. 
\item $|2,0,2\rangle$ if  $g_2<g_0,g_4$, \ie, for $c_1 >0$, $c_2>3c_1$. This situation may be accessed  with optical Feshbach resonances (see Appendix A for details).
\item $|2,0,4\rangle$ if $g_4<g_2, g_0$, \ie, for $c_1<0$ and $c_2>10c_1$; this case is hardly accessible experimentally. 
Using optical Feshbach resonance to achieve it, would lead to enormous losses (see Appendix A). 
\end{itemize}

In the case of singlets $|2,1,0\rangle$, the lattice ground state is non-degenerate in zeroth order in $t$, such that it does not have any effective dynamics. It does, however, have the first order correction to the wave function $|\Psi\rangle = (1 -(\hat H_0-E_0)^{-1}\hat H_t )\prod_i|2,1,0\rangle$. It has also 
the second order shift of the ground state energy $\delta E=-\langle\psi|\hat H_t (\hat H_0-E_0)\hat H_t|\psi\rangle$.   

\paragraph{\bf Pair Hamiltonian} To calculate the energy shifts to second order, we assume for simplicity 
that the energy spacing between eigenstates of $\hat{H}_0$ is sufficiently 
large compared to tunneling transitions to intermediate states and 
treat, therefore, these states as non-degenerate.
We shall consider here only the $|2,0,2\rangle$ single-site ground state, 
because the state $|2,0,4\rangle$ is hardly accessible experimentally.
The corresponding effective Hamiltonian admits a very large number of
virtual intermediate states and is, therefore, very complex to calculate.
We expect, however, that the physics for the cases $|2,0,2\rangle$
and $|2,0,4\rangle$ is similar. 

Starting  from $|2,0,2\rangle^{(i)}\otimes|2,0,2\rangle^{(j)}$, to second order in $t$
there are 26 intermediate virtual states spanning the five channels of total spin $S=0-4$:
\begin{equation}
\begin{split}
|1,0,2>^{(i)}\otimes|3,1,2>^{(j)}&\,\,\text{and}\,\,i\leftrightarrow j\,\,\,\,(S=0\ldots4),\\
|1,0,2>^{(i)}\otimes|3,0,0>^{(j)}&\,\,\text{and}\,\,i\leftrightarrow j\,\,\,\,(S=2=,\\
|1,0,2>^{(i)}\otimes|3,0,3>^{(j)}&\,\,\text{and}\,\,i\leftrightarrow j\,\,\,\,(S=1\ldots4),\\
|1,0,2>^{(i)}\otimes|3,0,4>^{(j)}&\,\,\text{and}\,\,i\leftrightarrow j\,\,\,\,(S=2\ldots4).
\end{split}\nonumber
\end{equation}
The corresponding energy shifts are:
\begin{eqnarray}
\epsilon_0&=&-t^2\frac{1}{7}\frac{\left(\frac{124}{35}\right)^2}{c_0+7/5 c_2},
\nonumber \\
\epsilon_1&=&-t^2
\left[\frac{1}{7}\frac{\left(\frac{22}{35}\right)^2}{c_0+7/5 c_2}
+\frac{2}{21}\frac{\left(\frac{144}{35}\right)^2}{c_0+3 c_1}\right],\nonumber\\
\epsilon_2&=&-t^2\left[\frac{1}{7}\frac{\left(\frac{6}{7}\right)^2}
{c_0+7/5 c_2}
+\frac{1}{15}\frac{\left(\frac{342}{49}\right)^2}{c_0-3 c_1}
+\frac{3 \left(\frac{20}{49}\right)^2}{c_0+3 c_1}\right.\nonumber\\
&+&\left.\frac{1}{11\cdot 35}\frac{\left(\frac{1124}{49}\right)^2}{c_0+7 c_1}\right],\nonumber \\
\epsilon_3&=&-t^2\left[\frac{1}{7}\frac{\left(\frac{8}{7}\right)^2}
{c_0+7/5 c_2}
+\frac{3}{7}\frac{1}{c_0+3 c_1}
+\frac{9}{11}\frac{\frac{21}{13}}{c_0+7 c_1}\right],\nonumber\\
\epsilon_4&=&-t^2\left[\frac{1}{7}\frac{\left(\frac{8}{7}\right)^2}
{c_0+7/5 c_2}+\frac{\left(\frac{18}{7}\right)^2}{c_0+3 c_1}
+\frac{5 \left(\frac{11}{7}\right)^2}{c_0+7 c_1}\right]\nonumber .
\end{eqnarray}
The full effective spin Hamiltonian reads
$\hat{H}= \sum_i \hat{H}_{0,i} + \sum_{<ij>} (\epsilon_0 \hat{P}_0^{(ij)} +
\epsilon_1 \hat{P}_1^{(ij)}+\epsilon_2 \hat{P}_2^{(ij)}+\epsilon_3 \hat{P}_3^{(ij)}
+\epsilon_4 \hat{P}_4^{(ij)})$. If scattering lengths are
changed through optical Feshbach resonances to have
$|2,0,2\rangle$ as the unperturbed single-site ground state, $\hat{H}$ has
a ferromagnetic lattice ground state 
$|\Psi\rangle=\bigotimes_i |2,0,2\rangle_2^{(i)}$ for $^{87}$Rb.
The energy shifts of the different spin channels are now mostly 
determined by the different contributions of the
intermediate states $|\nu\rangle$ and Clebsch--Gordon coefficients rather than 
by the differences in scattering length. 
As a result we do not expect that optical Feshbach resonances will allow such a precise and extensive
control of spin-spin interactions. We expect that in the $|2,0,4\rangle$ phase energy shifts exhibit a similar behavior
determined essentially  by the values of Clebsch--Gordon coefficients, although 
the  magnetic properties of this phase have yet to be calculated. 
Quite generally, we conjecture  that tuning spin-spin interactions \via optical Feshbach resonances in systems with two particles per site will be more difficult than for systems with one particle per site.

\section{IV. Ground state properties}

In this section we will study the ground state properties of systems with one or two particles per site, concentrating 
on the question whether the control over the scattering lengths 
can lead to the appearance of some more exotic phases (c.f. \cite{fisher}), assuming a simple
one dimensional chain or a two-dimensional square lattice. 
To this aim we use 
the approach developed recently by Wolf {\it et al.} \cite{wolf}, and search 
for such combinations of parameters for which we can represent the ground state of our Hamiltonian exactly (isolated exact ground states), or 
nearly exactly using matrix product states (MPS) \cite{werner} in 1D,  and projected entangled-pair states (PEPS) in 2D \cite{frankie}.

First we add to the bond Hamiltonian (\ref{1particleH})
a certain number of times the identity operator $\hat{I}^{(ij)}=\sum_S\hat{P}_S^{(ij)}$ on the bond, so that the Hamiltonian 
becomes positive definite, \ie,
\begin{equation}\label{1particH}
\hat{H}_I^{(ij)}=\sum_{S=0}^4\lambda_S \hat{P}_S^{(ij)},
\end{equation}
with all $\lambda_S$ being non-negative.

In the case of one atom per site (see Subsection A)
$\lambda_1=\lambda_3>0$ take the greatest values, whereas one of the 
$\lambda_S$ for $S=0,2,4$ can be set to zero. We will subsequently assume
that the control over the scattering lengths permits 
to choose which one is set to zero and to fix the magnitude of the others, even though this goes beyond the experimental 
feasibility; we will treat experimentally accessible cases with particular attention.  

In the case of two atoms per site, the values of $\lambda_S$ are 
essentially determined by the Clebsch-Gordan coefficients. They are
descending quite significantly  as $S$ increases,  with $\lambda_4=0$.
Thus an extensive control of the effective Hamiltonian using optical Feshbach resonances is
hardly  possible. Nevertheless, in Subsection B we explore some limiting cases by setting further
$\lambda_S$'s to zero.

\paragraph{\bf Seaching for isolated exact ground states} Our approach to search for exact (or, at least "variationally exact")  ground states can be described as follows. 
In 1D with periodic boundary conditions we seek for translationally invariant MPS of $N$ spins $F=2$, 
given by $|\Psi\rangle =\sum_{i_1,i_2,\ldots,i_N}{\rm Tr}( A^{i_1}A^{i_2}
\ldots A^{i_N})|i_1, i_2, \ldots, i_N\rangle$, where $i$'s enumerate the computational spin 2 basis, while $A$'s are 
$d\times d$ matrices, $d\ge1$. Let $R(ij)$ denote the range of the reduced density matrix 
$\rho_{ij}={\rm Tr}_{k\ne i,j}(|\Psi\rangle\langle \Psi|)$. This range is given by $R(ij)={\rm span}_M
\sum_{i,j}{\rm Tr}( MA^{i}A^{j})|i,j\rangle$, where the $M$'s are arbitrary $d\times d$  matrices. Let $K(ij)$ 
denote the kernel of the bond Hamiltonian (\ref{1particH}). Note that  if $R(ij)\subset K(ij)$,  then
$\hat H|\Psi\rangle=0$, and since $\hat H$ is non-negative, that implies that $|\Psi\rangle$ is a ground state. 
Similarly, in 2D with periodic boundary conditions we seek for translationally invariant PEPS of $N$ spins $F=2$ in 2D, 
given by $|\Psi\rangle =\sum_{{ i}_1,{ i}_2,\ldots,{ i}_N}{\rm Cont}( A^{{ i}_1}A^{{ i}_2}
\ldots A^{{ i}_N})|{ i}_1, { i}_2, \ldots, { i}_N\rangle$, where $i_k$'s enumerate the 
computational spin 2 basis of the 
$k$-th atom, and $k$ denotes coordinates in the 2D square lattice. 
This time  the tensors $A^{i_k}$ with four indices, all running from  $1$ to $d$, correspond to the 4 bonds touching the 
$k$th site. The contraction is over pairs of indices belonging to the same nearest neighbor bond. In this case 
 $R(ij)$ is given by $R(ij)={\rm span}_M
\sum_{i,j}{\rm Cont}( MA^{i}A^{j})|i,j\rangle$, where the $M$'s are arbitrary tensors with the 6 indices that are not contracted in the product of
$A^{i}A^{j}$. Alternatively, we may search for states that break the translational symmetry; antiferromagnetic N\'eel-like  ordering 
in 1D could  for instance 
correspond to MPS of the form
$|\Psi\rangle =\sum_{i_1,i_2,\ldots,i_N}{\rm Tr}( A^{i_1}B^{i_2}A^{i_3}
\ldots B^{i_N})|i_1, i_2, \ldots, i_N\rangle$ for even $N$.

\subsection{A. One atom per site}

\paragraph{\bf Mean field diagram} Before we proceed, it is worth to discuss the mean field phase diagram obtained under the assumption that the ground state is a product state, $|\Psi\rangle=|e,e,\ldots\rangle$ (see \cite{cyhPRA2000,barnett}). We are following here the most complete description of the phase diagram, provided recently in \cite{barnett}. We introduce the nematic tensor $Q_{ab}=\frac{1}{2}\ew{\hat F_a \hat F_b+\hat F_b \hat F_a}-\frac{1}{3}\delta_{ab}\ew{{\hat{\textbf{F}}}^2}$. There are 3 possible mean field (\ie, product) ground states, with $|e\rangle$ given up to $SO(3)$ rotations:
\begin{itemize}
\item Ferromagnetic state, $|e\rangle=(1,0,0,0,0)$; possesses only the $U(1)$ symmetry of rotations around the $z$--axis, and has maximal projection of the spin onto $z$ axis.
\item nematic state, which apart from $SO(3)$ rotation, has an additional $\eta$--degeneracy, $|e\rangle=(\sin(\eta)/\sqrt{2},0,
\cos(\eta),0,\sin(\eta)/\sqrt{2})$. This state is a MI version of the polar state in BEC; it has mean value of all components of the spin equal zero, but non vanishing singlet projection $\langle\rm singlet|e,e\rangle\ne 0$;
\item Tetrahedratic (cyclic) state, $|e\rangle=(1/\sqrt{3},0,0,\sqrt{2/3},0)$; this is a MI version of the cyclic state. The state may be uni- or biaxial, depending on whether the nematic  tensor does, or does not have a pair of degenerated eigenvalues; it has vanishing of both, of mean values of all of the spin components, and of the singlet projection. 
\end{itemize}
The phase  diagram is such that the system is in:  
\begin{itemize} 
\item a ferromagnetic state for $\lambda_4=0$, $\lambda_2,\lambda_0>0$, and for $\lambda_0=0$, provided $\lambda_2\ge 17\lambda_4/10$;
\item a nematic state  for $\lambda_0=0$, provided $3\lambda_4/10\lambda_2\le 17\lambda_4/10$;
\item a cyclic state for $\lambda_0=0$, provided $\lambda_2\le 3\lambda_4/10$, and for $\lambda_2=0$.
\end{itemize}

We will identify below the regimes of the phase diagram in which the mean field diagram is exact, or nearly exact.

\paragraph{\bf MPS reduce to mean field states} Since $\lambda_1=\lambda_3>0$ take the greatest values, 
the ground state of (\ref{1particH})
should belong to the symmetric subspace. For translationally invariant MPS this implies $[A^i,A^j]=0$, {\it i.e.}, the same matrix $K$ transforms 
$A^i$ and $A^j$ into the commuting Jordan forms. Generically, if all eigenvalues $\lambda_k^i, \lambda_k^j$  of $A^i,A^j$ are distinct, such MPS correspond to linear 
combinations of product states $|e_k,e_k,\ldots\rangle$, where $|e_k\rangle=\sum_i\lambda^i_k|i\rangle$. 
Since the Hamiltonian is a sum of nearest neighbor bond Hamiltonians, we have
$\sum_{k,k'}\langle e_k,e_k\ldots|\hat{H_I}|e_{k'},e_{k'}\ldots\rangle\propto\langle e_k|e_{k'}\rangle^{N-2}$ in a 1D-chain, and thus in the limit of an infinite chain
the ground states are equally well 
described by product states 
(that will typically break the rotational symmetry). This means in this case we expect mean field (product) states to provide a very good approximation of the ground states with translational symmetry. 
Let us analyze the different possible cases:

\noindent (A$_1$) For $\lambda_4=\lambda_2=\lambda_0=0$, all symmetric states are ground states, \ie, in particular all product states $|e,e\ldots\rangle$ 
with arbitrary $|e\rangle$.

\noindent (B$_1$) For $\lambda_4= \lambda_2=0,\ \lambda_0 >0$, 
the ground states $|e,e\ldots\rangle$ remind the cyclic states 
states of Ref.\ \cite{Demler} (\ie, they correspond to translationally but nor rotationally invariant product states), which now mix $S=2$ and $S=4$ contributions on each bond, and they have to fulfill the condition $\langle {\rm singlet}|e,e\rangle=0$. 
Denoting by $|e\rangle=(e_{2}, e_1,e_0, e_{-1}, e_{-2})$, this implies $e_0^2-2e_{1}e_{-1}+2e_{2}e_{-2}=0$. These states form a much greater class than the cyclic ones, since they may have non-vanishing (and even maximal) components of the spin. Interestingly, the transition between the cyclic phase for $\lambda_2=0$, and the ferromagnetic phase for $\lambda_4=0$, occurs via such states, \ie, at the transition point the degeneracy of the ground states manifold explodes. 

\noindent (C$_1$) For $\lambda_4=0$ and 
$\lambda_2,\,\lambda_0 >0$, the ground states are ferromagnetic states  $|2\rangle_{{\bf n}}|2\rangle_{{\bf n}}\cdots|2\rangle_{{\bf n}}$, corresponding to
a maximal projection of the local spin onto a given direction 
${\bf n=(\sin(\theta)\cos(\phi),\sin(\theta)\sin(\phi), \cos(\theta))}$.
Such vectors for $F=2$ may be parametrized (in the basis of $\hat{\bf F}_{\bf n}$ with descending
$m_F$) as
$\propto(z^{-2},2z^{-1},\sqrt6,2z,z^2)$ with $z=|z|e^{i\phi}, \,\, 
|z|\in(- \infty,\infty)$. It should be stressed that ferromagnetic states are {\it exact} ground state in the entire 
part of the phase diagram whenever $\lambda_4=0$. 

\begin{figure}[t]
\includegraphics[width=0.8\linewidth]{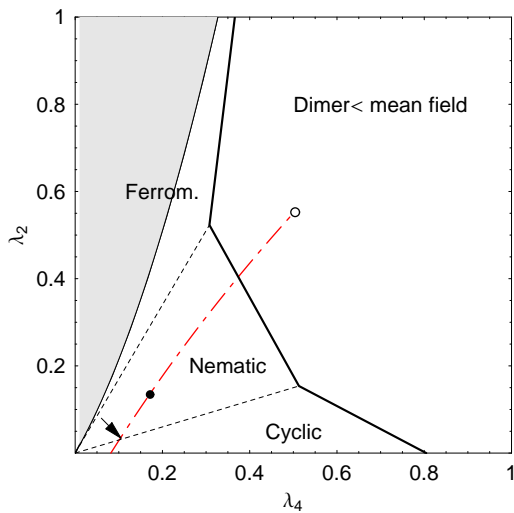}
\caption{Sketch of the phase diagram for the (D$_1$) case,
obtained by applying the variational principle 
in the $\lambda_2, \,\lambda_4$ phase space 
(for $\lambda_0=0$) to mean field, N\'eel, and dimer states with one atom per site. 
The scale is set by letting $\lambda_1=\lambda_3=1$.
N\'eel-type states are never favorable over nematic states.  The {\it ferromagnetic} region (gray) was obtained  
numerically by imaginary time evolution of MPS, and comparing 
the result from runs with $d=1$  and $d=5$ in a chain of $50$ sites with open boundary conditions \cite{frankie1}. Of course,  on the line $(\lambda_4=0,\lambda_2)$ ferromagnetic states give always ground states. 
Dashed lines
indicate the regions where the type of mean field state with lowest energy changes qualitatively
(see text for more details).
The red (dashed-dotted) line gives the values of $(\lambda_2,\lambda_4)$ which can be obtained by changing the
spin-independent scattering length $\bar c_0=(3 \bar g_4+4 \bar g_2)/7$ of $^{87}$Rb through optical Feshbach resonances.
The arrow gives the values for unchanged $\bar c_0$,
black and white circles indicate a change of $\bar c_0$ of $10$\% and $100$\%, respectively.
}
\label{fig:phases}
\end{figure}

\noindent (D$_1$) For $\lambda_0=0$ and
$\lambda_4,\,\lambda_2 >0$, the ground states apparently favor 
antiferromagnetic order. This, however, can be misleading, if $\lambda_4\ll\lambda_2$. In that case, as the mean field diagram suggests, the ferromagnetic order might prevail. We have applied in 1D a more general variational approach, going beyond mean field. We have looked for ground states by 
applying the variational principle to mean field (product)  states $|e,e\ldots\rangle$, 
N\'eel-type states $|e,f,e,f\ldots\rangle$, 
and  valence bond solid states with singlet states for distinct pairs (dimers) of neighboring atoms and translational dimer symmetry.
For the mean field case as discussed earlier
the energy is either minimized by the ferromagnetic state $|e\rangle=|2\rangle_{\bf n}$ (for $\lambda_2\ge 17\lambda_4/10$), by a nematic state $|e\rangle=|0\rangle_{\bf n}$ (for $3\lambda_4/10\le\lambda_2\le 17\lambda_4/10$; in this case the state is a combination of total spin $0,\,2$ and $4$), or, for $\lambda_2\leq3\lambda_4/10$, by a cyclic state, $|e\rangle=(e_2,e_1,e_0,e_{-1},e_{-2})$ with $e_0=1/\sqrt2,\,e_2=-e_{-2}=1/2,\,e_1=e_{-1}=0$. 
Imposing N\'eel order with $\langle e|f\rangle\neq1$ always results in a larger energy, as $\lambda_{1,3}>\lambda_{2,4}$, and the overlap
with the singlet can be maximized already by restricting to product states. On the other hand, for the dimer state 
the energy per bond is given by $\frac12{\rm Tr}(H_I\frac1{25}\mathbbm 1\otimes\mathbbm 1)$. 
This results in the phase diagram shown in Fig.~\ref{fig:phases}, somewhat analogous to the results obtained by 
Yip \cite{yip1}. The red line depicts values of $\lambda_2,\,\lambda_4$ experimentally accessible through modifications of scattering lengths. We have applied MPS code to  
search numerically for the exact ground states using the method of \cite{frankie1}. We confirmed that in the shaded region 
in Fig.~\ref{fig:phases} there is indeed a ferromagnetic ground state. We have also studied the ground state at the experimentally accessible line, and found indications of nematic and dimer order in the phase diagram, see Fig.~\ref{fig:phases}. We expect that in 2D, in addition a possible ground state could be formed from resonating valence bond states \cite{fisher}, and we are planning to apply the 2D PEPS methods to investigate this question. 
\begin{figure}[t]
\includegraphics[width=0.77\linewidth]{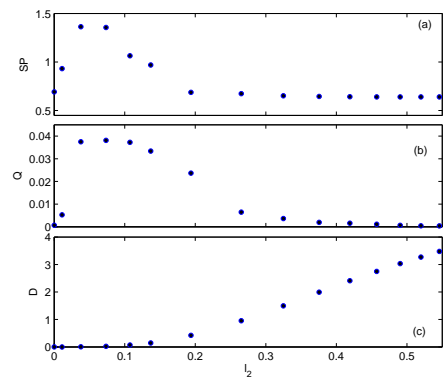}
\caption{
Analysis of the ground states obtained variationally from an MPS ansatz for the case (D$_1$), for combinations
of $\lambda_2$ and $\lambda_4$ lying on the dotted-dashed line of Fig.~\ref{fig:phases}.
(a) Singlet projection $SP=\sum_i{\rm tr}(\hat P_0\rho_i\otimes\rho_i)$,
where $\rho_i$ is the reduced density matrix of site $i$,
(b) nematic order parameter $Q=\max_{\omega}Q_{\Omega}=\max_{\Omega}\sum_i[(\vec{n}_{\Omega}\vec{S}_i)^2-2]/N$ ($\vec{n}_{\Omega}$ is the unit vector pointing in direction $\Omega$), and
(c) dimerization $D=|\ew{\hat P_0^{N/2,N/2+1}}-\ew{\hat P_0^{N/2-1,N/2}}$.
(Values obtained for open chains of $16$ sites, with $d$ up to $30$.
}
\label{fig:phases2}
\end{figure}

\noindent (E$_1$) For $\lambda_2=0$ and $\lambda_4,\,\lambda_0 >0$, 
as in the (D$_1$) case, mean field cyclic states are favorable over N\'eel states. 
We have compared them  variationally to the analogues of the dimer states in the present case, \ie, 
configurations which have a state with total spin $S=2$ on distinct bonds.
We call these state {\it para--dimers}. Now the situation is quite different
from the dimerized states discussed in (D$_1$), as the states on the bond
are not unique and states with different total spin projection $M_{S=2}$ can form
superpositions. In the subspace of states having a para-dimer on each second bond,
the Hamiltonian can be written as an effective interaction between neighboring
para--dimers, $(H'_{\rm eff})_{(ij)}=\sum_{S'=0}^4\lambda_{S'}(P')^{(ij)}_{S'}$,
where $i,\,j$ now enumerate the para--dimers, and $(P')_{S'}$ projects onto the subspace
of two para-dimers with total spin $S'$. For $\lambda_2=0,\,\lambda_0,\,\lambda_4>0$,
always $\lambda'_2<\lambda'_0,\,\lambda'_4$, and the optimal superposition of para--dimers
with different projections is again the cyclic combination. On the other hand, given that
$\lambda'_2$ is the lowest coefficient, combining neighboring para--dimers to states with
total spin $S'=2$ might lead to even lower energies. On this level again states with
different total $z$-projection $M_{S'}$ can be combined, and it turns out that again
the cyclic combination minimizes the energy. Comparing the energies of cyclic product
states, cyclic states of para--dimers, and cyclic combinations of "para-dimerized" para-dimers,
the phase diagram shown in Fig.~\ref{fig:phases2} is obtained.


%
%
\begin{figure}[t]
\includegraphics[width=0.8\linewidth]{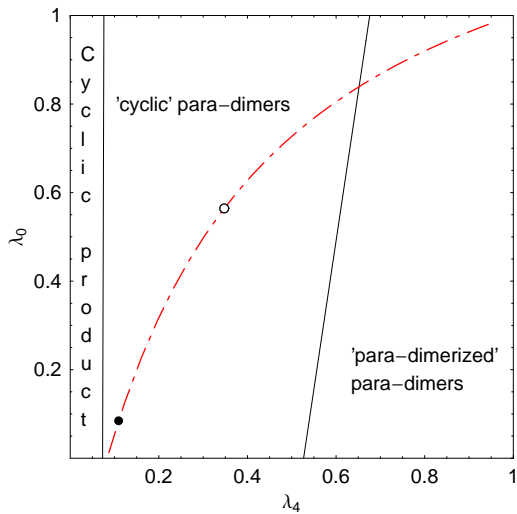}
\caption{Sketch of the phase diagram for the case (E$_1$) case, \ie, when $\lambda_2$ is the lowest coefficient.
As in Fig.~\ref{fig:phases}, the Hamiltonian is shifted and rescaled, such that $\lambda_2=0$,
$0\leq\lambda_0,\,\lambda_4\leq1$, and $\lambda_1=\lambda_3=1$. The lines give the boundaries
obtained from variationally comparing product states (where cyclic states are always optimal) to
product states of para--dimers (where again the cyclic combination gives minimal energy), and
combinations of neighboring para--dimers with total spin $S'=2$. The red (dotted--dashed) line indicates
the combinations of $(\lambda_4,\lambda_0)$ which can be obtained by changing $\bar c_0$
through an optical Feshbach resonance. Black and white circles indicate a change of $\bar c_0$
of $-10\%$ and $-50\%$, respectively. 
}
\label{fig:phases2}
\end{figure}

\subsection{B. Two atoms per site}

We discuss here cases where the values of the $\lambda_S$ are in descending order,
determined essentially by the Clebsch-Gordan coefficients. In such a situation
only a ferromagnetic-like order in the ground state is possible, but now the states do not neccessarily have to be  of the product form, especially if $\lambda_1=0$, or $\lambda_1=\lambda_3=0$.  We explore the following limiting cases:

\noindent (A$_2$) For all $\lambda_S=0$ except $\lambda_0>0$, 
the ground states, when reduced to neighboring sites, are either of the type
$|2\rangle_{{\bf n}}|e\rangle$ (or $|e\rangle|2\rangle_{{\bf n}}$) with  
$|e\rangle=(e_2, e_1, e_0,e_{-1},0)$ in the ${\bf S_n}$ basis) or in the 
form $|1\rangle_{{\bf n}}|\tilde e\rangle$ with 
$|\tilde e\rangle=(e_2, e_1, e_0,0,0)$ (or $|\tilde e\rangle|1\rangle_{{\bf n}}$).

\noindent (B$_2$)  Similarly, for $\lambda_2=\lambda_3= \lambda_4=0$
and $\lambda_0,\lambda_1>0$, ground states correspond to product states which, 
when reduced to neighboring sites, have either the form 
$|2\rangle_{{\bf n}}|e\rangle$ (or $|e\rangle|2\rangle_{{\bf n}}$) with $|e\rangle=(e_2, e_1, e_0,0,0)$ written in the ${\bf S_n}$ basis, or the form $|1\rangle_{{\bf n}}|1\rangle_{{\bf n}}$.

\noindent  (C$_2$) For $\lambda_3= \lambda_4=0$  and all other 
$\lambda_S >0$, the ground states are product states formed 
by the vectors $|2\rangle_{{\bf n}}$, and $|1\rangle_{{\bf n}}$, 
with the constraint that there are not two $|1\rangle_{{\bf n}}$ 
states in the neighboring sites.

\noindent (D$_2$) for $\lambda_4=0$ and all other 
$\lambda_S >0$, the ground states are as in the case (C$_1$) before, \ie,
they are of the form 
$|2\rangle_{{\bf n}}|2\rangle_{{\bf n}}\cdots|2\rangle_{{\bf n}}$.

\section{V. $F=3/2$  Fermi-Hubbard Hamiltonian}

\paragraph{\bf The system} Let us now turn to the discussion of the spin-$3/2$ (in this and the next section)
and -$5/2$ (in Secs.~VII, VIII) Fermi lattice gases. 
 The total wavefunction of the fermions has to be anti-symmetric, implying that 
the spin of two colliding fermions can only be even. Interaction for two
fermions with spin $F$  in the $s$-wave channel can be written in the form 
\begin{equation}
  \hat{V} = \bar g_0\hat{P}_0+\bar g_2\hat{P}_2+...+\bar g_{2F-1}\hat{P}_{2F-1},
\end{equation}
where \(\hat{P}_S\) is the projection operator on the subspace with total spin $S$ and \(\bar g_S\) is 
the interaction strength, which depends on the scattering length (\(a_S\)) 
\(\bar g_S=\frac{4\pi \hbar^2a_S}{m}\).

\paragraph{\bf On-site Hamiltonian} For two spin-$3/2$ particles with anti-symmetric
spin-wavefunction, we can use the identities 
\(\hat{I} = \hat{P}_{0}+\hat{P}_2\) and 
\(\hat {\bf F}_1\cdot \hat {\bf F}_2 = \gamma_0 \hat{P}_0+\gamma_2 \hat{P}_2\) 
to express the interaction in the form \(\hat{V} = \bar c_0\hat{1} + \bar c_2 \hat {\bf F}_1\cdot \hat {\bf F}_2\).
Here, \(\hat{\bf F}_i\) is the spin operator of the particle $i$ and 
\(\gamma _n = [n(n+1)-2F(F+1)]/2\). The total Hamiltonian in the limit of vanishing tunneling ($t=0$)
for $F=3/2$ is a sum of single-site Hamiltonians of the form (omitting site indices)
\begin{equation}
\begin{split}
\label{ground_hamiltonian}
  \hat H_0 = &\frac{c_0}{2} 
  \sum_{\alpha\beta}\hat a_{\alpha} ^\dagger \hat a_{\beta} ^\dagger
  \hat a_{\beta} \hat a_{\alpha} + \\ 
  &\frac{c_2}{2}
  \sum_{\alpha\beta\gamma\sigma}\hat a_{\alpha} ^\dagger \hat a_{\beta} ^\dagger
  (F_{3/2})_{\alpha\gamma}(F_{3/2})_{\beta\sigma} \hat a_{\sigma} \hat a_{\gamma},
\end{split}
\end{equation}
where \(c_0=(-g_0+5g_2)/4\) and \(c_2=(-g_0+g_2)/3\), and symbols without
bars are related to those with bars in the same way as in the bosonic sections. In the summation,
greek letters are spin indices.

The eigenvalues of the Hamiltonian \(\hat H_0\) read
$$
E^0(N,\bar{F})=\frac12c_0N(N-1) + \frac12c_2(\bar{F}(\bar{F}+1)-\frac{15}4N),
$$
where \(N\) is the number
of particles per site and \(\bar{F}\) is the total on-site spin. 
As there are four accessible states per site, corresponding to the
spin projections, the maximal number of particles is $N=4$. 
The  energies for different values of $N$ and $\bar F$ are listed below:
\begin{equation}
\begin{split}
&E^0(N=1,\bar{F}=3/2) = 0, \\
&E^0(N=2,\bar{F}=2) = c_0-\frac{3c_2}{4}=g_2, \\
&E^0(N=2,\bar{F}=0) = c_0-\frac{15c_2}{4} =g_0, \\
&E^0(N=3,\bar{F}=3/2) = 3c_0 - \frac{15c_2}{4} =(g_0+5g_2)/2, \\
&E^0(N=4,\bar{F}=0) = 6c_0 - \frac{15c_2}{2}=g_0+5g_2.
\end{split}
\end{equation}

\begin{figure}[th]
\includegraphics[width=0.9\linewidth]{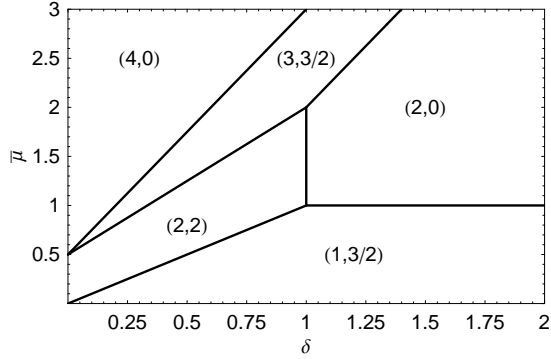}
\caption{Ground states for $F=3/2$ for the case of no tunneling, in the space
of $\delta=g_2/g_0$ and $\bar \mu=\mu/g_0$. The phases
are labeled by $(N,\bar F)$, where $N$ denotes the number
of particles per site and $\bar F$ their total spin.
}
\label{fig:phases3half}
\end{figure}

\paragraph{\bf Phases at $t=0$} 
The actual ground state (GS) in the case of vanishing tunneling is determined by comparing the
(Gibbs potential) energies $G=E_0(N,\bar F)-\mu N$ of the above
listed states. The resulting phase diagram is already quite complex
and depends on the values of \(c_0\) and \(c_2\), or better to say $g_0, g_2$,
in a nontrivial way (see Fig.~\ref{fig:phases3half}). Denoting the states
by $(N,\bar F)$, and writing $\delta=g_2/g_0$ and $\bar\mu=\mu/g_0$, we obtain
\begin{itemize}
\item (0,0) is the GS for $\bar \mu<0$;
\item (1,3/2) is the GS for $0<\bar\mu<1$, and $\bar\mu<\delta$;
\item (2,0) is the GS for $1<\bar\mu<(5\delta-1)/2$, and $\delta>1$;
\item (2,2) is the GS for  $\delta<1$, and $\delta<\bar\mu<(3\delta+1)/2$;
\item (3,3/2) is the GS for ${\rm max}[(3\delta+1)/2, (5\delta-1)/2]<\bar\mu<
(5\delta+1)/2$;
\item (4,0) is the GS for $(5\delta+1)/2<\bar\mu$.
\end{itemize}

\section{VI. Effective Hamiltonian for $F=3/2$}

We follow here the same lines as in bosonic part. We assume weak tunneling, such that
we can use perturbation theory to calculate an effective spin--spin Hamiltonian.
The tunneling Hamiltonian
\(H_t= -t\sum_{<ij>,\sigma}( a_{i\sigma}^\dagger a_{j\sigma}+a_{j\sigma}^\dagger a_{i\sigma})\)
as before involves nearest neighbors only and conserves the total spin $S$ as well as its
$z$-component $M_S$. We apply Eq.~(\ref{eq:eps_s}) to calculate energy shifts
to second order in $t$, limiting the calculations to one value of $M_S$, which we will
choose to be the largest.


\subsection{A. One atom  per site}

We consider first the case of one particle in each lattice site. For $t=0$, the ground state of the 
nearest neighbour pair with total spin $S$ and maximal total $z$-projection can be written as
\begin{equation}
\label{ground_1_particle}
  |SM_S=S\rangle _{i,j} = \sum_{m_1,m_2} \langle FF \, m_1 m_2 | SM_S=S \rangle \,
  a_{im_1}^\dagger a_{jm_2}^\dagger |\Omega\rangle,
\end{equation}
where $F$ is a spin of a fermion (\(m_1\), \(m_2\) are the $z$-components of this spin), 
and $i$ and $j$ are lattice indices.

Possible intermediate states are those having two particles on one site (say $i$), and
no particles on the other (say $j$). Two-particle states with total on-site spin $\bar F$
and maximal projection $M_{\bar F}=\bar F$ read
\begin{eqnarray}
  |\nu\rangle_{ij} &=& \ket{\bar F,M_{\bar F}=\bar F}_{ij}\nonumber\\
  &=&\frac{1}{\sqrt{2}}\sum_{m_1 \neq m_2}
  \langle FF \, m_1 m_2 | \bar F,m_{\bar F}=\bar F \rangle \,
  a_{im_1}^\dagger a_{im_2}^\dagger |\Omega\rangle.\label{m_1_particle}
\end{eqnarray}
From anti-commutation relations for fermions and properties
of Clebsch-Gordon coefficients, it follows that $\bar F$ has to be even.
The energy shifts thus are simply
\begin{equation}
\begin{split}
\epsilon _S &= \, -\frac{4t^2}{g_S}, \qquad\text{(total spin) $S$ is even},\\
\epsilon _S &= \, 0,\qquad\qquad\text{S is odd}.
\end{split}
\end{equation}
This in fact is independent of the spin $F$.
The explicit expressions for  $F=3/2$ can be written
\begin{equation}
\begin{split}
  \epsilon _2 =& -\frac{4t^2}{g_2} = -\frac{4t^2}{c_0-3c_2/4} \\
  \epsilon _0 =& -\frac{4t^2}{g_0} = -\frac{4t^2}{c_0-15c_2/4}\\
  \epsilon _3 =& \, \epsilon _1 = 0,
\end{split}\label{eqn:shifts1ppsF}
\end{equation}
and the effective Hamiltonian is $\hat H_I^{(ij)}=\epsilon_2\hat P_2^{(ij)}+\epsilon_4\hat P_4^{(ij)}$.
This result is the same as discussed in Ref. \cite{tu}. The effective spin model has
several particularly interesting limits. In particular, when $g_0=g_2$ the model
has  a $SU(4)$ symmetry, and is integrable via Bethe ansatz in 1D \cite{sutherland};
its ground state is a spin singlet with gapless excitations.

\subsection{B. Two atoms per site}

We now consider the case of 
two spin-$3/2$ fermions on each lattice site, so that the total spin on-site may take values 0 or 2.
We express two-site states in the form
\(|S,\bar{F}\rangle\), where $S$ is the total spin of the two sites and \(\bar{F}\) 
is the total on-site spin. Again we can limit calculations to the maximal total $z$-component $M_S=S$.
We write the ground state in the same form as in Eq. \eqref{ground_1_particle}
\begin{eqnarray}
\label{ground_2_particle}
  |S,\bar{F}\rangle _{i,k} &=& 
  \frac{1}{2}\!\!\!\sum_{\substack{m_1,m_2\\n_1,n_2,\bar{M},\bar{N}}}\!\!\!
  \langle \bar F\bar F\bar M\bar N | SM_S=S\rangle
  \langle FF \, m_1 m_2 | \bar{F}\bar{M} \rangle\cdot\nonumber\\
  & & \cdot\langle FF \, n_1 n_2 | \bar{F}\bar{N} \rangle \,
  a_{im_1}^\dagger a_{im_2}^\dagger a_{kn_1}^\dagger a_{kn_2}^\dagger |\Omega\rangle,
\end{eqnarray}
where \(\bar{M},\bar{N}\) are the $z$-components of the total spins of sites $i,\,k$, respectively.
Intermediate states $\ket{\nu}$ now have three fermions in site $i$ (which is equivalent
to having a spin-3/2 hole) and one in the neighboring site $k$. As in Eq. \eqref{ground_2_particle},
we explicitly give the intermediate states with total spin $S$ and maximal total $z$-component $M_S$.
Such virtual states exist only for integer $S=0\ldots3$. 
\begin{equation}
\begin{split}
\label{m_2_particle}
  |S=3\rangle =& 
  a_{i\frac32}^\dagger a_{i\frac12}^\dagger a_{i-\frac12}^\dagger a_{k\frac32}^\dagger\ket{\Omega} \\
  |S=2\rangle =& \frac{1}{\sqrt{2}} \left( 
  a_{i\frac32}^\dagger a_{i\frac12}^\dagger a_{i-\frac12}^\dagger a_{k\frac12}^\dagger - 
  a_{i\frac32}^\dagger a_{i\frac12}^\dagger a_{i-\frac32}^\dagger a_{k\frac32}^\dagger \right)\ket{\Omega} \\
  |S=1\rangle =& \frac{1}{\sqrt{10}} \big( \sqrt{3}
  a_{i\frac32}^\dagger a_{i\frac12}^\dagger a_{i-\frac12}^\dagger a_{k-\frac12}^\dagger - \\
  &2
  a_{i\frac32}^\dagger a_{i\frac12}^\dagger a_{i-\frac32}^\dagger a_{k\frac12}^\dagger + \sqrt{3}
  a_{i\frac32}^\dagger a_{i-\frac12}^\dagger a_{i-\frac32}^\dagger a_{k\frac32}^\dagger \big)\ket{\Omega} \\
  |S=0\rangle =& \frac{1}{2} \big(
  a_{i\frac32}^\dagger a_{i\frac12}^\dagger a_{i-\frac12}^\dagger a_{k-\frac32}^\dagger -
  a_{i\frac32}^\dagger a_{i\frac12}^\dagger a_{i-\frac32}^\dagger a_{k-\frac12}^\dagger + \\
 &a_{i\frac32}^\dagger a_{i-\frac12}^\dagger a_{i-\frac32}^\dagger a_{k\frac12}^\dagger -
  a_{i\frac12}^\dagger a_{i-\frac12}^\dagger a_{i-\frac32}^\dagger a_{k\frac32}^\dagger \big)\ket{\Omega}.
\end{split}
\end{equation}
Using those states, the energy shifts to second order can be calculated.
If the total on-site spin is $\bar F=0$, \ie, the on-site states are singlets, then those do not interact
in second order perturbation theory, but there is a second order shift which amounts to  
$\epsilon_{0} = -2t^2/(c_0+15c_2/4)$ per bond.
If the atoms on-site form a composite with spin $2$, then the energy shifts are as follows: 
\begin{equation}
\begin{split}
  \epsilon _{4} =\epsilon _{2} =& 0, \\
  \epsilon _{3} =\epsilon _{1} =& -\frac{4t^2}{c_0 - 9c_2/4}, \\
  \epsilon _{0} =& -\frac{10t^2}{c_0 - 9c_2/4}. 
  \end{split}
\end{equation}
The above results agree with those obtained recently  
in Ref.~\cite{tu}. 

\subsection{C. Three atoms per site}

For three particles per lattice site, possible two-site states are similar to the case of 
one atom per site, Eq.~\eqref{m_1_particle}, because states of three atoms per site can equivalently be written
as a single hole in the filled Fermi sea. Intermediate states now have four particles on one site.
Because of Pauli's principle there is only one such state, namely the filled Fermi sea
\(a_{i,3/2}^\dagger a_{i,1/2}^\dagger a_{i,-1/2}^\dagger a_{i,-3/2}^\dagger|\Omega\rangle\). 
It is thus clear that the energy shifts have to be as in Eq.~(\ref{eqn:shifts1ppsF}):
\begin{equation}
\begin{split}
  \epsilon _2 =& -\frac{4t^2}{g_2} = -\frac{4t^2}{c_0-3c_2/4}, \\
  \epsilon _0 =& -\frac{4t^2}{g_0} = -\frac{4t^2}{c_0-15c_2/4},\\
  \epsilon _3 =& \, \epsilon _1 = 0.
\end{split}
\end{equation}

Obviously, for four particles per site the on-site ground state is the filled Fermi sea.
This is an exact eigenstate of the full Hamiltonian, as no tunnelings are possible
in this state.

\section{VII. $F=5/2$  Fermi-Hubbard Hamiltonian}\label{sec:7}

The theory for insulating states of a spin-$5/2$ gas with $1$ or $2$ atoms
per lattice site is essentially the same as in the case of spin-$3/2$ particles, so  
we will comment only the basic differences.

\paragraph{\bf The system and on-site states}
Now the two-particle s-wave interaction can be written in the form
\begin{equation}
\hat{V} = \bar g_0\hat{P}_0+\bar g_2\hat{P}_2+\bar g_4\hat{P}_4.
\end{equation}
or, using the identity operator \(\hat{I}\) and spin operators
\(\hat {\bf F}_i\),
\begin{equation}
\hat{V}=\bar c_0 \hat{1} + \bar c_1 (\hat {\bf F}_i\cdot \hat {\bf F}_j) + \bar c_2 \hat{P}_0,
\end{equation}
where \(\bar c_0=(5\bar g_2+23\bar g_4)/28\), \(\bar c_1=(-\bar g_2+\bar g_4)/7\), and 
\(\bar c_2=(7\bar g_0-10\bar g_2+3\bar g_4)/7\).
The singlet projection operator \(\hat{P}_0\) can be represented {\it via} creation and 
annihilation operators as
\begin{eqnarray}
\hat{P}_0 &=& \hat{A}^\dagger \hat{A}, \\
\hat{A} &=& -\frac{1}{\sqrt{3}} 
\left(
\hat a_{\frac52}\hat a_{-\frac52} - \hat a_{\frac32}\hat a_{-\frac32} + \hat a_{\frac12}\hat a_{-\frac12}
\right).
\end{eqnarray}
The on-site Hamiltonian attains then  a similar form as in the case of spin-3/2 with
an  additional term \(c_2\hat{P}_0\) (and relations between $\bar c_i$ and $c_i$ as before).
The energies for different numbers of fermions per site $N$ and total
on-site spin $\bar F$ are listed below:
\begin{equation}
\begin{split}
&E^0(N=1,\bar{F}=5/2)= 0 \\
&E^0(N=2,\bar F=4) = c_0 +\frac{5}{4}c_1 \\
&E^0(N=2,\bar F=2) = c_0 -\frac{23}{4}c_1 \\
&E^0(N=2,\bar F=0) = c_0 -\frac{35}{4}c_1 + c_2 \\
&E^0(N=3,\bar{F}=9/2) = 3c_0 -\frac{3}{4}c_1 \\
&E^0(N=3,\bar{F}=5/2) = 3c_0 -\frac{35}{4}c_1 + \frac{2}{3}c_2 \\
&E^0(N=3,\bar{F}=3/2) = 3c_0 -\frac{45}{4}c_1,
\end{split}
\end{equation}
\begin{figure}[th]
\includegraphics[width=0.9\linewidth]{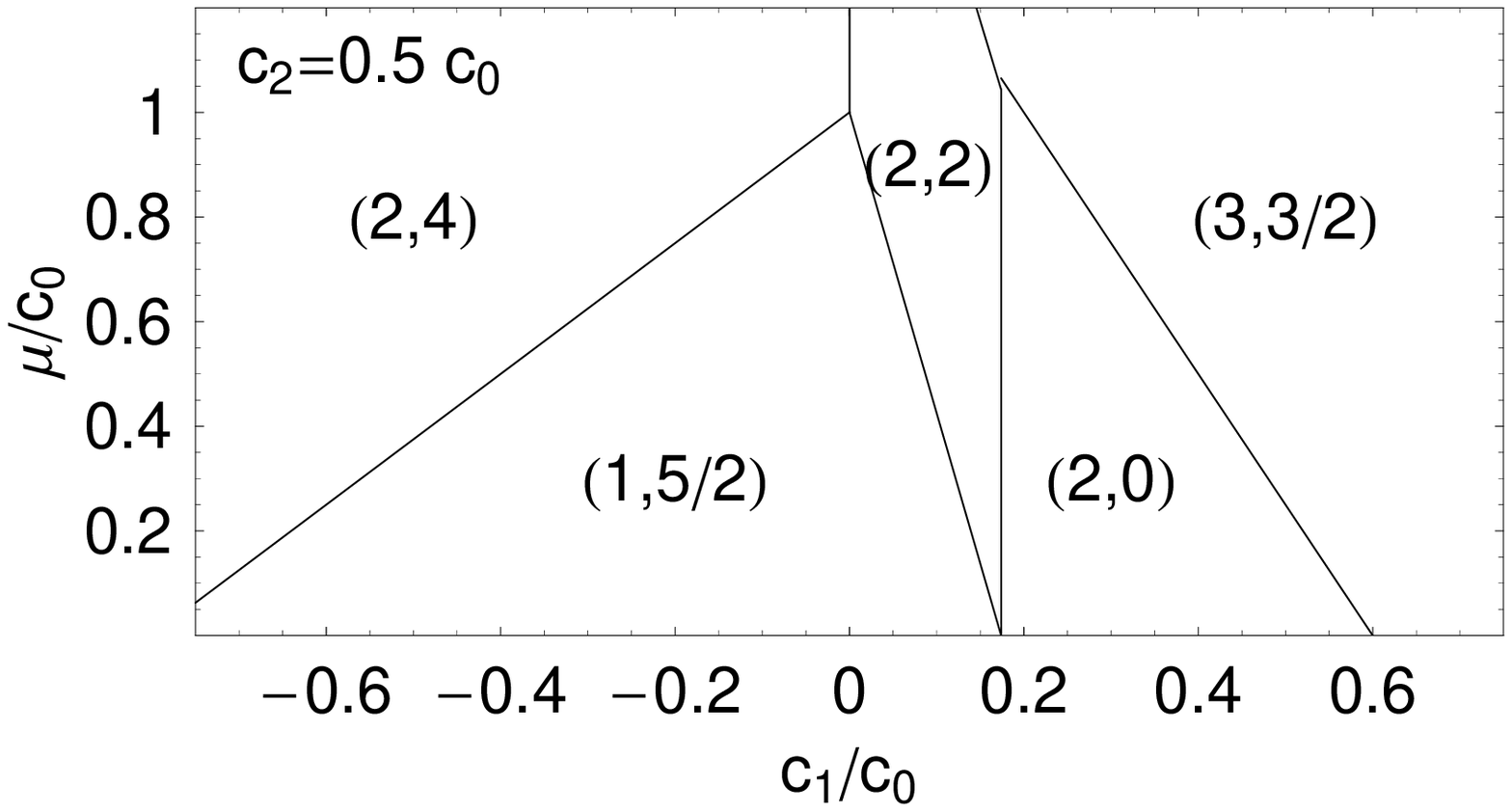}\\\vspace{0.3cm}\includegraphics[width=0.9\linewidth]{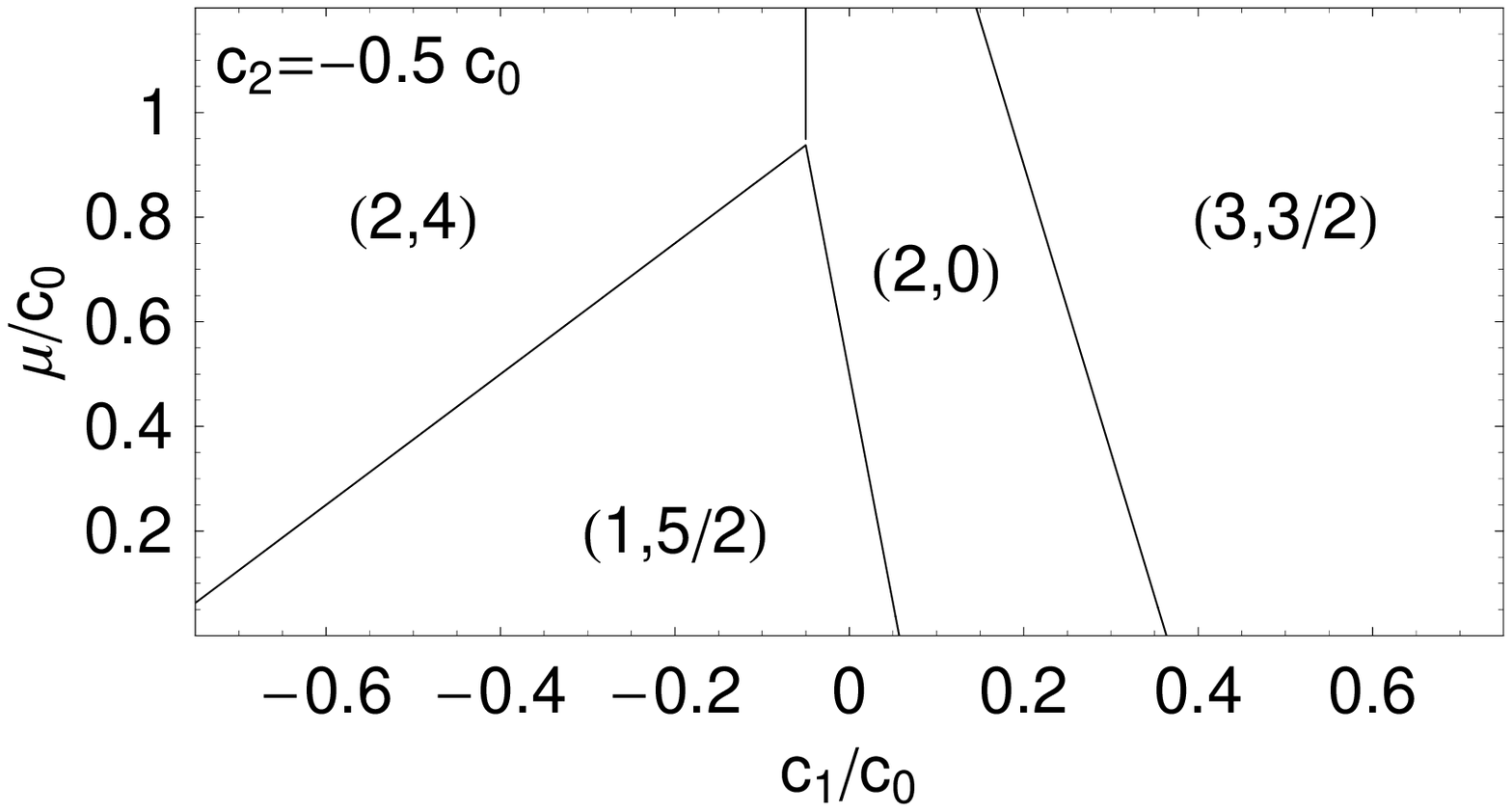}
\caption{Ground states for $F=5/2$ for the case of no tunneling, in the space
of $c_1/c_0$ and $\bar \mu=\mu/g_0$ for $c_2/c_0=1/2$ (top) and $c_2/c_0=-1/2$ (bottom),
taking into account states of up to $3$ particles.
The phases are labeled by $(N,\bar F)$, where $N$ denotes the number
of particles per site and $\bar F$ their total spin.
}
\label{fig:phases5half}
\end{figure}

\paragraph{\bf Phases at $t=0$}
The actual ground states are determined by minimizing the
Gibbs energy $G=E_0(N,\bar F)-\mu N$. The resulting phase diagram 
is three-dimensional and quite complex, as it depends on the values of 
\(c_0\)  $c_1$, and \(c_2\) (or better to say $g_0,\,g_1$, and $g_2$) in
a highly nontrivial way. The ground states are plotted in the space of
$\mu/c_0$ and $c_1/c_0$ for two values of $c_2$ in Fig.~\ref{fig:phases5half}.

\section{VIII. Effective Hamiltonian for $F=5/2$}


\subsection{A.  One atom per site}

\begin{figure}[t]
\includegraphics[width=0.95\linewidth]{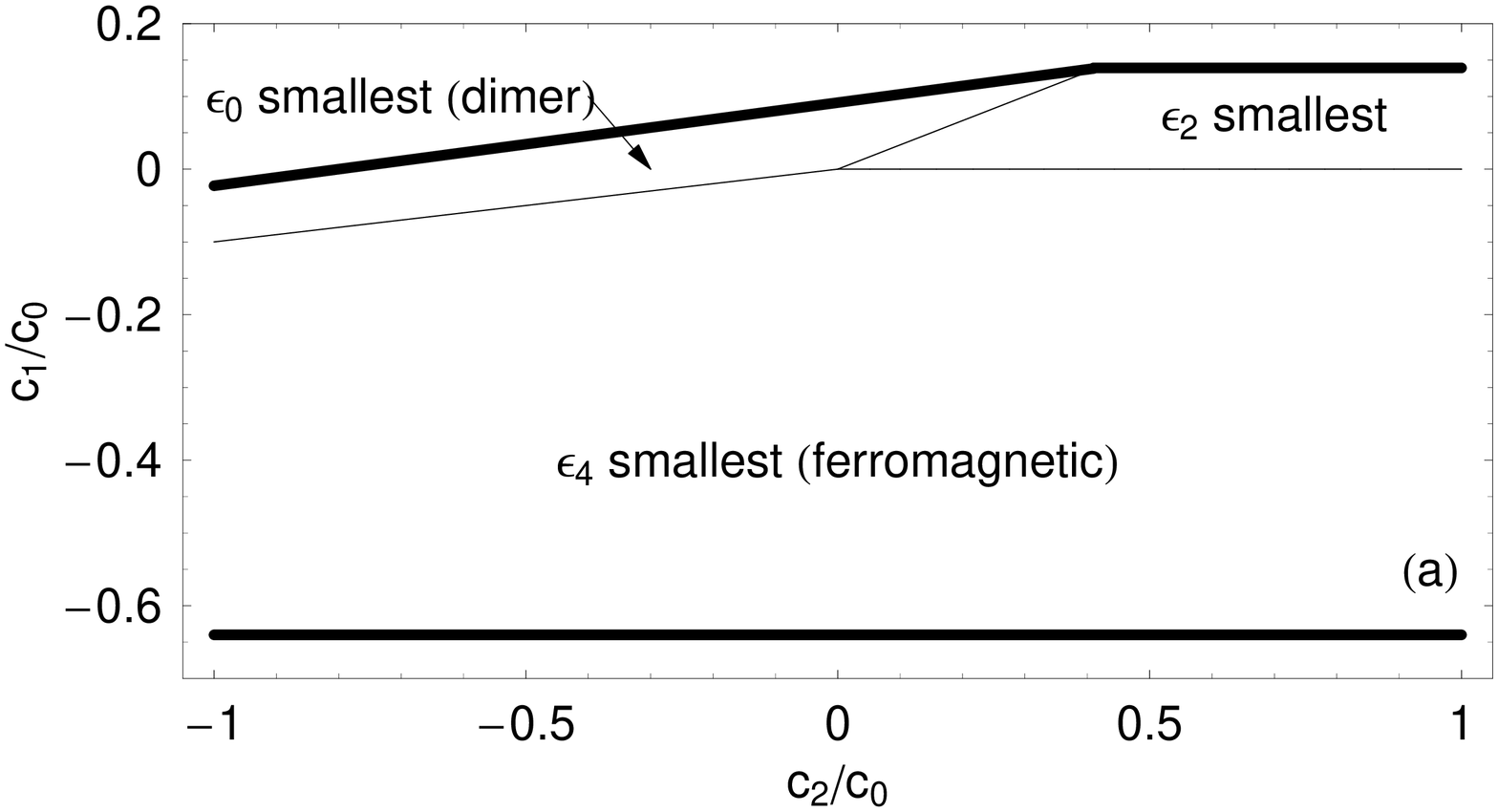}\\\vspace{0.2cm}
\includegraphics[width=0.9\linewidth]{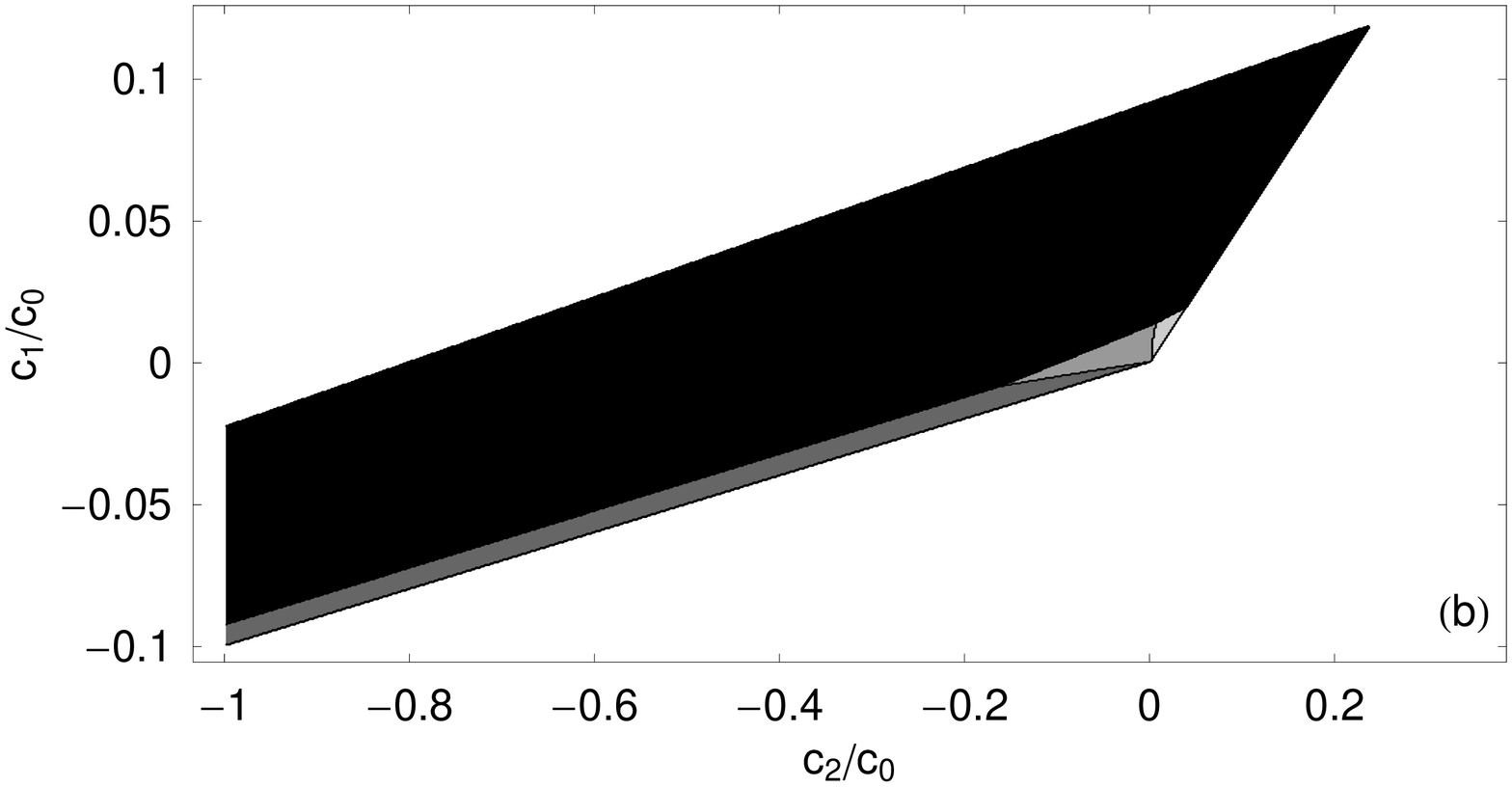}
\caption{
(a) Classification of the effective Hamiltonians which can be obtained for $F=5/2$ and a single particle per site
(see Eq.~(\ref{eq:521aps}))
in the $(c_1/c_0,\,c_2/c_0)$ space (for $\mu/c_0=0.2$). Thick lines indicate the borders of the region where
the $t=0$ ground state has a single particle per site, thin lines indicate the borders between
different regimes of the corresponding effective Hamiltonian. When $\epsilon_4$ is 
lowest, then the ground state is ferromagnetic. The area with $\epsilon_0<\epsilon_2,\epsilon_4$,
is shown in more detail in (b). The different regions are obtained from a variational ansatz similar to
Sec.~IV, case (D$_1$), comparing dimer with different mean field type states.
Lowest energy states according to this ansatz are mostly dimerized (black).
The ferromagnetic (dark grey), nematic (grey), and cyclic (light grey) regions are much smaller.
}
\label{fig:phases5half1pps}
\end{figure}

The effective Hamiltonian to second order in the tunneling amplitude $t$
has the form \(\hat H_{\rm I}^{(ij)} = 
\epsilon _0 \hat{P}_0^{(ij)} + \epsilon _2 \hat{P}_2^{(ij)}+ 
\epsilon _4 \hat{P}_4^{(ij)}\), where
\begin{equation}
\begin{split}
  \epsilon _{4} =& -\frac{4t^2}{g_4} = -\frac{4t^2}{c_0 +\frac{5}{4}c_1}, \\
  \epsilon _{2} =& -\frac{4t^2}{g_2} = -\frac{4t^2}{c_0 -\frac{23}{4}c_1}, \\
  \epsilon _{0} =& -\frac{4t^2}{g_0} = -\frac{4t^2}{c_0 -\frac{35}{4}c_1 + c_2}.\\
\end{split}\label{eq:521aps}
\end{equation}
As usually $\epsilon_4$ is smallest, the ground states in this case are mostly
ferromagnetic, as can be seen from Fig.~\ref{fig:phases5half1pps} (a). There is
however a region where $\epsilon_0<\epsilon_2,\epsilon_4$.  In this case the
variational approach followed in Sec.~IV, case ($D_1$) shows that within this 
region again dimerized as well as ferromagnetic, nematic, or cyclic phases can be realized
(see Fig.~\ref{fig:phases5half1pps} (b)).%

\subsection{B. Two atoms per site}

On-site ground states with two fermions per site can have total spin $0$, $2$, or $4$.
Tunneling carries over these states into those with three atoms on one site, and
one on the neighboring site. The state of three particles on lattice site $i$
with total spin $\bar F$ and $z$-projection $\bar M$ can be written in the form
\begin{eqnarray}
\label{3-particle}
|\bar F\bar M\rangle \propto \!\!\!\!\!\sum_{\substack{m_1,m_2,m_3\\\bar F_2,\bar M_2}}\!\!\!\!\!\!\!\!&&
\langle FF \, m_1 m_2 | \bar F_2\bar M_2 \rangle
\langle \bar F_2F \, \bar M_2 m_2 | \bar F \bar M \rangle \! \times\nonumber\\
&&\times \,a_{i,m_1}^\dagger a_{i,m_2}^\dagger a_{i,m_3}^\dagger\ket{\Omega}.
\end{eqnarray}
As the virtual state necessarily has to be anti-symmetric, there are three
different possibilities for $\bar F$, namely $9/2$, $5/2$, or $3/2$.
Let us now consider ground states with different on-site spin
$\bar F$ separately. The case $\bar F=0$ leads only
to a ground state energy shift, but not to interesting dynamics
in second order perturbation theory. 

\paragraph{\bf On-site ground states with spin $4$}
In this case
nearest--neighbor pairs can form total spin from $S=0$ to $S=8$. The effective
Hamiltonian in second order perturbation theory can be written in the form
$H_{\rm I}^{(ij)} = \sum_S \epsilon _S \hat{P}_S^{(ij)}$, with
\begin{equation}
\begin{split}
\epsilon _8 =& 0, \\
\epsilon _7 =& -\frac{4t^2}{c_0-\frac{13}4c_1}, \\
\epsilon _6 =& 0, \\
\epsilon _5 =& -\frac{\frac{15}{7}t^2}{c_0-\frac{13}4c_1} -
  \frac{\frac{13}7t^2}{c_0-\frac{45}4c_1+\frac23c_2}, \\
\epsilon _4 =& -\frac{\frac{7865}{2058}t^2}{c_0-\frac{13}4c_1} -
  \frac{\frac{143}{98}t^2}{c_0-\frac{45}4c_1+\frac23c_2} -
  \frac{\frac{572}{1029}t^2}{c_0-\frac{55}4c_1}, \\  
\epsilon _3 =& -\frac{\frac{1875}{686}t^2}{c_0-\frac{13}4c_1} -
  \frac{\frac{11}{98}t^2}{c_0-\frac{45}4c_1+\frac23c_2} -
  \frac{\frac{396}{343}t^2}{c_0-\frac{55}4c_1}, \\
\epsilon _2 =& -\frac{\frac{297}{343}t^2}{c_0-\frac{13}4c_1} -
  \frac{\frac{33}{49}t^2}{c_0-\frac{45}4c_1+\frac23c_2} -
  \frac{\frac{396}{343}t^2}{c_0-\frac{55}4c_1}, \\
\epsilon _1 =& 
  -\frac{\frac{24}7t^2}{c_0-\frac{45}4c_1+\frac23c_2} -
  \frac{\frac47t^2}{c_0-\frac{55}4c_1}, \\
\epsilon _0 =& -\frac{6t^2}{c_0-\frac{45}4c_1+\frac23c_2}.
\end{split}
\end{equation}
Tuning $c_1/c_0$ and $c_2/c_0$, typically either $\epsilon_8$ and $\epsilon_6$, $\epsilon_7$, or $\epsilon_4$ are
the smallest coefficients, such that, though often ferromagnetic ground states are realized,
also models preferring anti-ferromagnetic order are possible.

\paragraph{\bf On-site ground states  with spin $2$}
Finally, we consider the case of on-site ground states with spin $2$,
where
\begin{equation}
\begin{split}
\epsilon _4 =& 
  -\frac{\frac{825}{686}t^2}{c_0+\frac{43}4c_1} -
  \frac{\frac{45}{98}t^2}{c_0+\frac{11}4c_1+\frac23c_2} -
  \frac{\frac{60}{343}t^2}{c_0+\frac14c_1}, \\
\epsilon _3 =& 
  -\frac{\frac{825}{686}t^2}{c_0+\frac{43}4c_1} -
  \frac{\frac{81}{98}t^2}{c_0+\frac{11}4c_1+\frac23c_2}-
  \frac{\frac{676}{343}t^2}{c_0+\frac14c_1}, \\
\epsilon _2 =& 
  - \frac{\frac{150}{343}t^2}{c_0+\frac{43}4c_1} -
  \frac{\frac{50}{147}t^2}{c_0+\frac{11}4c_1+\frac23c_2} -
  \frac{\frac{200}{343}t^2}{c_0+\frac14c_1}, \\
\epsilon _1 =& 
  - \frac{\frac47t^2}{c_0+\frac{11}4c_1+\frac23c_2} -
  \frac{\frac{24}7t^2}{c_0+\frac14c_1}, \\
\epsilon _0 =&
  -\frac{\frac{10}3t^2}{c_0+\frac{11}4c_1+\frac23c_2}.
\end{split}
\end{equation}
Now the effective Hamiltonian is
\(H_{\rm eff} = \sum_i H_{0,i}+\sum_{<ij>}
\epsilon _0 \hat{P}_0^{ij}+\epsilon _1 \hat{P}_1^{ij}+
\epsilon _2 \hat{P}_2^{ij}+\epsilon _3 \hat{P}_3^{ij}+\epsilon _4 \hat{P}_4^{ij}\).
Typically, either $\epsilon_1$ or $\epsilon_0$ are the smallest coefficients,
such that realizable spin models usually have ground states preferring anti-ferromagnetic order over ferromagnetic
one.

\section{IX. Conclusions}

Summarizing, in the first part of the article we have analyzed the different Mott insulating phases of repulsive spinor
$F=2$ bosons confined in optical lattices at low temperatures. 
We have discussed two experimentally relevant cases with either 
one or two atoms per lattice site. Our analysis shows that in the case 
of a single atom per lattice site, the spin-spin couplings and, therefore, 
the magnetic properties of the system can be precisely manipulated 
using optical Feshbach resonances. We have explored the quantum phase diagram 
for such a case using variational and numerical techniques. 
On the other hand, the manipulation of the magnetic properties 
of a ground state with two atoms per lattice site becomes  
much harder to achieve. In this last case,  
the spin-spin interactions present in the effective Hamiltonian 
couple many different virtual states. Since (at zero magnetic field) the scattering 
lengths $a_S$ are very similar, the 
couplings depend very strongly on the corresponding Clebsch-Gordan coefficients.  In this respect, spinor condensates with non-alkaline atoms \cite{pfau} which present large differences of their scattering lengths $a_S$ could display stronger spin-spin interactions effects \cite{ho2005}. As it was pointed out to us by L.~Santos, in the limit when $|t/g_0|$ is very small, the effective spin-spin interactions might involve magnetic dipole-dipole interactions. Still, for Rubidium the 
effective model here is valid for a certain range of $|t/g_0|$ for which the magnetic dipole moment can be neglected. On the other hand, tuning the system into a range where dipole-dipole interactions are important, they might offer an additional knob to control the effective Hamiltonian.

In the second part we have performed a similar analysis of $F=3/2$ and $F=5/2$ gases. Also there, while the phases with one 
fermion (or one fermionic hole, i.e., $2F$ fermions) per lattice site can be more easily controlled with optical Feshbach resonances, the physics of phases with 2 or 4 atoms is controlled by the Clebsch-Gordon coefficients. The latter situation
might still lead to various spin models preferring either ferromagnetically or anti-ferromagnetically ordered ground states. In this context it would be particularly appealing to realize (e.g., by the above mentioned dipole-dipole interactions) Hamiltonians with dominant contribution of $\hat P_3$-terms. Such Hamiltonians admit AKLT-like gapped ground states in 2D in the honeycomb lattice \cite{future}.

\section{Acknowledgments}

We thank  K. Bongs, J.I. Cirac, E. Demler, J. Eschner, M. Guilleumas, M. Mitchell, J. Mur-Petit, A. Polls, E. Polzik, L. Santos, and K. Sengstock for discussions. 
We acknowledge support from Deutsche Forschungsgemeinschaft (SFB 407, SPP 1116, GK 282, 436 POL), EU IP Programme ``SCALA'', 
European Science Foundation PESC QUDEDIS,
and MEC (Spanish Government) under contracts FIS 2005-04627, FIS 2005-01369, EX2005-0830, Consolider/Ingenio 2010 "QOIT".
\L.\,Z.~and M.J.L.~thank the ICFO -- Institut de Ci\`{e}ncies Fot\`{o}niques for hospitality.

\appendix

\section{Appendix: Optical Feshbach resonance for $F=2$ $^{87}$ Rb atoms}

Optical modifications of scattering length, or in other words optical Feshbach resonances (OFR) were proposed in Ref. \cite{fedichev}, and carefully analyzed theoretically in a series of papers by J.L. Bohn and P.S. Julienne \cite{bohn}. These authors have pointed out that OFRs are inevitably associated with spontaneous emission losses, since molecular states used for OFR cannot be to far from the photoassociation resonance. For these reason, changes of natural scattering length of  $^{87}$Rb, which itself is of order 100 a.u., by more than 10 a.u. were considered to be unrealistic. These prediction have been confirmed in the recent experiment of R. Grimms group \cite{grimm}. 

For the present investigations this implies that only limited changes of scattering length are possible. Note that since OFR takes place far from the nucleus,  where the excited state potential has a dipol form, $-C_3/R^3$, 
One can only modify in this way a spin independent part of the scattering,
T hat is $(3a_4+4a_2)/7$. That means that $a_0$ and $4a_4-3a_2$ remain unchanged under OFR.   	

The most accurate value of $a_2=(91.28\pm 0.2)$a.u. \cite{verhaar}. I. Blochs group has studied collisionally driven spin dynamics of $^{87}$Rb in Mott insulator regime in an optical lattice \cite{blochas}, and measured very precisely scattering length differences. From these measurements we obtain: $a_0=87.77\pm 0.4$a.u., and $a_4= 97.23 \pm 0.2$a.u. That implies that
$4a_4-3a_2=115.08$a.u. Assuming that one may modify spin independent scattering by 10\%, we get $(3a_4+4a_2)/7=93.83 \pm 9.4$. We see that $a_4$ may vary   
roughly as $a_4=97.23\pm 7.9$a.u., along the line  $4a_4-3a_2=115.08$a.u.

This estimate has important consequences: there are no feasible OFR, that could  reach the regime $a_4<a_0$. This is the reason, why reaching the regime of Mott state with 2 atoms per site, and total spin 4 (\ie, $|2,0,4\rangle$) is hardly possible with OFRs.

\section{MPS and PEPS: A Quantum Information approach to strongly correlated systems}

The density matrix renormalisation group (DMRG)\cite{White92,Schollwoeck05} is a variational 
method that has had an enormous success in describing ground states 
of some strongly interacting 1D systems with rather 
modest computational effort. 
The underlying philosophy of all DMRG oriented algorithms 
is that many body systems can be treated almost "exactly" if one is
able to truncate the full Hilbert space by removing the degrees of freedom that
are not involved neither in the ground state, nor in the 
dynamical evolution of the system. The difficulty and glory of the method
relies on how reliable the truncation is done.
Very recently \cite{Vidal03a,Vidal03b,frankie1,orus04}, 
Quantum Information Theory has provided a new perspective on the following questions: 
(i) how to perform an efficient truncation of the Hilbert space, 
(ii) which quantum systems can be efficiently simulated, 
(iii) how to simulate dynamical Hamiltonian and dissipative evolutions of strongly correlated systems, 
(iv) how and when DMRG-oriented methods can be 
implemented to investigate ground states of 2D and
3D  systems;
(v) how classical concepts like correlation length, which diverge
on the critical points is linked to the entanglement\cite{Verstraete04a}, etc.
In perhaps in the simplest version QI approach reduces to variational methods based on matrix product states, or more general  
projected entangled pair states
In general, this Quantum Information approach, apart from being extremely simple to implement, is  very efficient 
for  strongly correlated systems and id already 
shedding a new light in our understanding of many body physics. Among recent successes of QI approach are: 
efficient codes for periodic boundary conditions \cite{frankie1}, simulations of finite $T$
 and dissipative systems \cite{Verstraete04c}, renormalization algorithms for Quantum-Many Body Systems in two and higher dimensions \cite{frankie},  understanding of the role of entanglement in quantum phase transitions \cite{Verstraete04a},
efficient evaluation of partition functions of frustrated and inhomogeneous spin systems \cite{murg},
 and spectra of excited states \cite{diego}, studies of quantum impurity modesl \cite{uli}, simulation of critical 
\cite{vidalcrit}, and infinite-size \cite{vidalinf} quantum lattice systems in 1D, MPS representations of Laughlin wave functions \cite{jillast}, simulating quantum adiabatic aproach to hard NP-problems \cite{jil2}, MPS based image compressions \cite{jil3},  just to name few.

\end{document}